\documentclass[12pt]{article}

\usepackage[textheight=22.5truecm,textwidth=16.8truecm,top=2.95truecm,left=2.15truecm]{geometry}
\usepackage{
amsmath,
amsthm,
amssymb,
ascmac,
bm,
tikz,
braket,
mathrsfs,
graphicx,
color,
hyperref,
cite,
titlesec,
caption
}

\hypersetup{colorlinks,bookmarksopen,bookmarksnumbered,citecolor=black,linkcolor=black,linktocpage,pdfstartview=FitH,urlcolor=black}

\numberwithin{equation}{section}

{\makeatletter
\long\def\@makefntext#1{\parindent 1em\noindent 
\@hangfrom{\hbox to 1.8em{\hss$^{\@thefnmark}$}}#1}
\makeatother}

\def\fnum@figure{\textbf{\figurename\nobreakspace\thefigure}}
\def\fnum@table{\textbf{\tablename\nobreakspace\thetable}}
\long\def\@makecaption#1#2{%
  \vskip\abovecaptionskip
  \sbox\@tempboxa{\small #1. #2}%
  \ifdim \wd\@tempboxa >\hsize
    \small #1. #2\par
  \else
    \global \@minipagefalse
    \hb@xt@\hsize{\hfil\box\@tempboxa\hfil}%
  \fi
  \vskip\belowcaptionskip}
\renewcommand{\l}[0]{\left}
\renewcommand{\r}[0]{\right}
\renewcommand{\d}[0]{\mathrm{d}}

\allowdisplaybreaks

    \title{\hfill\parbox{3cm}{\normalsize KUNS-2944}\\[12pt]
Spacetime-emergent ring\\
toward 
tabletop quantum gravity experiments
}
\author{
Koji Hashimoto\footnote{koji@scphys.kyoto-u.ac.jp},\quad
Daichi Takeda\footnote{takedai@gauge.scphys.kyoto-u.ac.jp}, \\
Koichiro Tanaka\footnote{kochan@scphys.kyoto-u.ac.jp},\quad and\quad
Shingo Yonezawa\footnote{yonezawa.shingo.3m@kyoto-u.ac.jp}\\[12pt]
 \textit{ Department of Physics, Kyoto University, Kyoto 606-8502, Japan}
}
\date{}

\begin{document}
\maketitle
\begin{abstract}
	We propose a way to discover, in tabletop experiments, spacetime-emergent materials, that is, materials holographically dual to higher-dimensional quantum gravity systems under the AdS/CFT correspondence.
	The emergence of the holographic spacetime is verified by a mathematical imaging transform of the response function on the material. We consider theories on a 1-dimensional ring-shaped material, and compute the response to a scalar source locally put at a point on the ring. 
	When the theory on the material has a gravity dual, 
	the imaging in the low temperature phase exhibits a distinct difference from the ordinary materials: the spacetime-emergent material can look into the holographically emergent higher-dimensional curved spacetime and provides an image as if a wave had propagated there.
		Therefore the image is an experimental signature 
	of the spacetime emergence.
	We also estimate temperature, ring size and source frequency usable in experiments, with an example of a quantum critical material, TlCuCl$_3$. 
\end{abstract}

\newpage

\tableofcontents

\section{Introduction}\label{sec: introduction}
In contrast to the variety of theoretical scenarios considered so far, quantum gravity has not yet been tested experimentally. 
There are two major directions for quantum gravity experiments: direct measurement of quantum gravity corrections in our universe, and studying materials as emergent gravity systems.
Since the former approach requires the energy around the Planck scale, which is a severe obstacle,
it is reasonable to pioneer the latter possibility, tabletop experiments for quantum emergent gravity.

Holography, the AdS/CFT (anti-de Sitter spacetime/conformal field theory)  correspondence \cite{Maldacena:1997re}, offers the progress for that. 
Non-holographic examples of materials which exhibit effective gravitational phenomena include the acoustic black holes \cite{pelat2020acoustic} and the type I\hspace{-1pt}I topological material \cite{soluyanov2015type}, while holography is rather an established framework which can provide a duality to truly quantum gravity, not just to the classical part of the gravity or just curved spacetimes.
In this paper, to build a foundation for quantum gravity tests, we propose a way to discover ring-shaped \textit{spacetime-emergent materials} (SEMs), materials equivalent to higher-dimensional quantum gravity systems under the AdS$_{3}$/CFT$_{2}$ correspondence.\footnote{
The subscripts denote their spacetime dimensions.
A short introduction of the AdS/CFT will be given soon after.
}
We demonstrate theoretical calculations of the response function on the ring, and find that optical imaging transformation, which is performed mathematically, helps discriminate materials letting 3-dimensional gravity spacetimes be emergent holographically.
That is, we claim that, by imaging the response signal on the ring, SEMs can be discovered in laboratory tabletop experiments. 
The discovery of an SEM leads directly to the quantum gravity tests, as it amounts to a technology to create our own universe of the micro- or nano-scale size, which is the playground for the experimental study of Hawking radiation, information loss problem of black holes, and even the birth of the universe.

In the remaining of this section, we will position our work, following the history of the AdS/CFT in a way friendly to readers who are not familiar with the holography.

\begin{figure}
	\centering
	
	\includegraphics[width = 5cm]{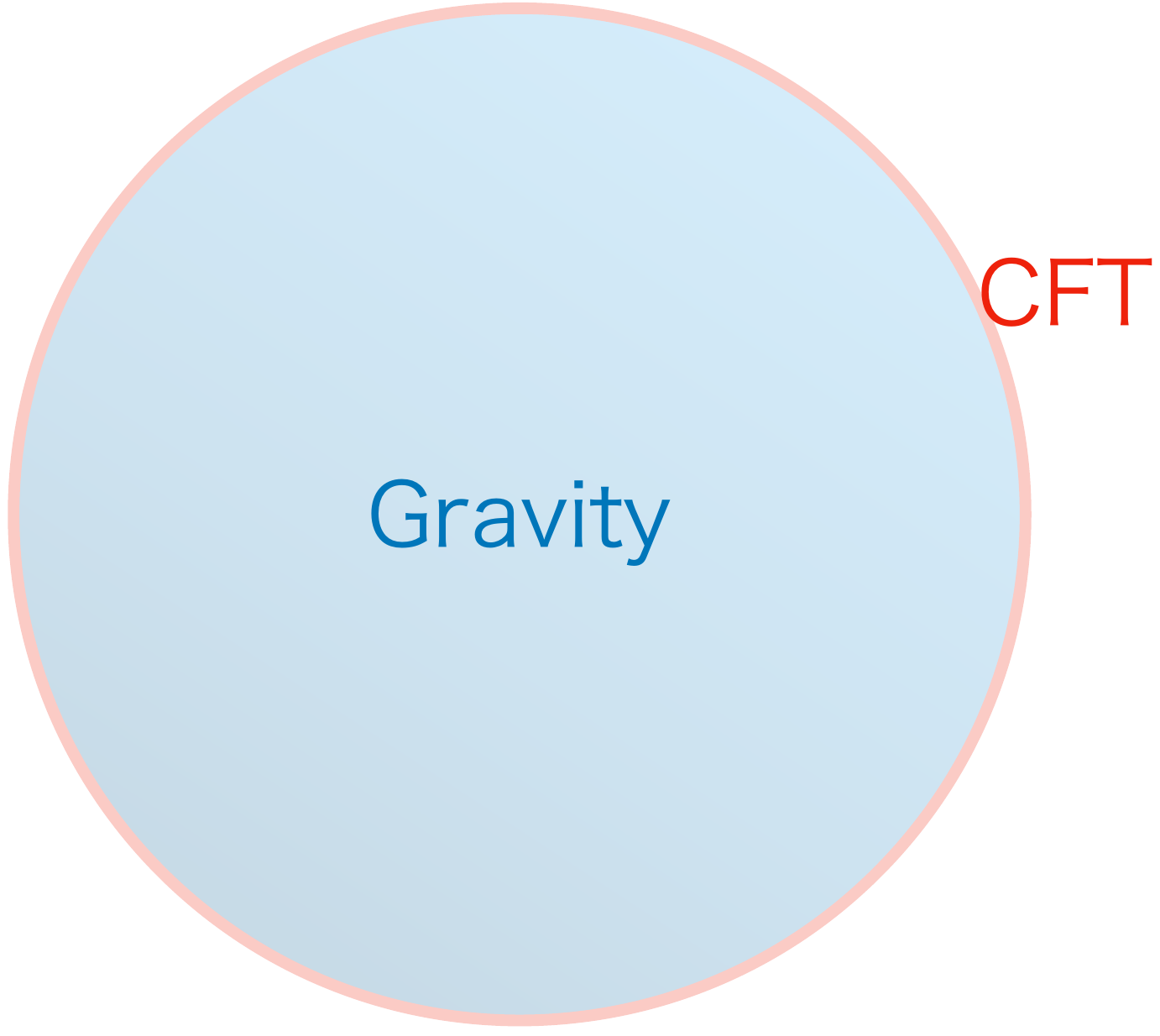}
	\caption{
	The illustration of the AdS/CFT correspondence.
	The bulk spacetime hosts a gravity system, while the boundary spacetime hosts the material CFT. The CFT is equivalent to the gravity.
	The time coordinate 
	is not depicted here.
	}
	\label{fig: AdS/CFT}
\end{figure}

CFT (conformal field theory) is a non-gravitational relativistic quantum field theory which has the conformal symmetry (and thus it is scale-free), and AdS (anti-de Sitter spacetime) is a maximally symmetric spacetime which solves the Einstein equation with a negative cosmological constant.
The AdS/CFT correspondence, dubbed as the holographic duality, claims that a certain CFT is equivalent to some higher-dimensional system containing gravitational degrees of freedom on the AdS spacetime background.
Given that such a CFT$_d$ is defined on a $(d-1)$-dimensional sphere $\mathbb S^{d-1}$ (which is a compact space with the unique scale, the radius of the sphere\footnote{When $d=2$, the sphere is a ring. We call this as CFT$_d$ because the spacetime is a $d$-dimensional $\mathbb R \times \mathbb S^{d-1}$ where $\mathbb R$ denotes the time coordinate.}), then the dual gravity theory is to be defined inside the sphere with the time coordinate shared, the AdS$_{d+1}$ spacetime (see Fig.\ref{fig: AdS/CFT}).
Hence where the CFT is defined is called the boundary, and the AdS side is the bulk.
As this duality originates in string theory, quantum nature of the gravity is already a part of the formulation, and especially 
the bulk Planck scale is not necessarily so large as that of our universe.
This is why it is important to find materials following the AdS/CFT.

Since at zero temperature, CFT is scale-free unless compactified, it is expected that, near quantum critical points, there could be materials well approximated by some CFTs.
Thus, the AdS/CFT suggests the existence of SEMs.
Studies related to this are called AdS/CMP (condensed matter physics) \cite{Karch:2007pd}.
In this field so far, variety of gravity models that mimic phase diagrams of some condensed matters have been built, for example as one of the most successful stories, the holographic superconductor \cite{Hartnoll:2008vx,Hartnoll:2008kx}.
However, just by examining aspects universal among certain materials, one cannot conclude for which material among them a spacetime is emergent, and what kind of spacetime emerges.\footnote{
In constructing holographic CMP models, one may also think of comparing the theoretical spectra with experiments, but such a bottom-up construction of the gravity model having the same spectrum is almost equivalent to finding a new AdS/CFT example, so it is definitely a difficult problem.}
Is there any reasonable method which can judge whether a material that we bring to a laboratory is an SEM or not?

In the AdS/CFT, a black hole can appear in the bulk when the boundary material is at nonzero temperature. 
It was shown theoretically in \cite{Hashimoto:2018okj, Hashimoto:2019jmw} that visual image of holographic black holes can be read from observables of SEMs (see also \cite{Kaku:2021xqp, Zeng:2022woh}).
Here the \textit{holographic black hole} means a black hole virtually appearing in the emergent spacetime.
In the papers, the Einstein ring of the Schwarzschild-AdS$_4$ black hole was visualized, just as in our universe the image of the supermassive black hole in M87 was observed by the Event Horizon Telescope.
Since the photographs of the Einstein rings are naturally interpreted as a result of an optical wave which had propagated in the emergent curved spacetime, the imaging of holographic black holes can directly catch the spacetime emergence.
Along this novel possibility of detection of the spacetime emergence, we make a further step for realistic experiments --- spacetime-emergent rings. In fact, we notice that, so long as we deal with AdS$_{4}$/CFT$_{3}$ in which the material shape is a 2-dimensional sphere $\mathbb S^{2}$, we cannot apply the strategy to real experiments, because it is in general technically difficult to process a quantum material into a stable sphere.
Thus we in this paper propose to use $\mathbb S^{1}$ instead --- ring-shaped materials, which are relatively easy to make in laboratories. With one-dimension lowered, the correspondence is now AdS$_{3}$/CFT$_{2}$.

We will also consider the well-known gravitational phase transition between the black hole phase (high temperature) and the AdS soliton phase (low temperature). In the transition the bulk geometry changes from the one with a black hole to the one without but with an energy gap.
This corresponds to the conductor/insulator transition of the boundary CFT in the context of the AdS/CMP.\footnote{The CFT is a scale-free theory, thus has no phase-transition scale. The only scale of this CFT on the ring is the ring circumference $a$. Therefore the phase transition occurs at the temperature $T=1/a$. This rather peculiar transition temperature of the SEM is in contrast to the ordinary phase transition in materials. }
There are mainly two reasons for us to deal with the transition.
First, since we are planning to realize experiments, 
to spot a candidate material we need to compare the transition with the one for the material.
The second reason, which is more critical, is that for the case of the SEM ring the bulk of the high temperature phase is not probed well by the imaging.
All the bulk light rays shot from the boundary are absorbed by the BTZ black hole (the holographic black hole in AdS$_{3}$), and never come back to the boundary again, meaning that the bulk information is lost.
Therefore, as proven in \cite{Hashimoto:2018okj}, the straightforward application of the imaging strategy to the BTZ fails.
On the other hand, we will later find that the image in the AdS soliton phase (the low temperature phase) is so bright only at the antipodal point of the source location, while not when we observe it at other points on the ring.
This characteristic signal of the image of the response is the sign of the emergent spacetime --- 
when we discover in experiments a material whose image has that characteristics, we can claim that the material is an SEM.\footnote{Although there may be variety of possible modified SEM models, the feature is physically so firm that we expect they should still exhibit the feature, as discussed in section \ref{sec: discussion}.}
To clarify the universality of the feature of the signal, we also compare the result with that of a non-SEM theory, i.e., a model of materials which does not give any dual gravity.

The organization of this paper is as follows.
In section 2, we introduce the setup and our strategy more in detail, but in a way friendly to people who are not familiar with or not interested in technical aspects of the holography.
In section 3, the response function of an ordinary material (non-SEM) and that of an SEM are computed respectively.
The results are mathematically processed by the imaging method and shown in section 4.
In section 5, raising TlCuCl$_3$ as a candidate, we estimate parameters needed to judge in experiments if it is an SEM.
Section 6 is devoted to a summary and discussions.


\section{Setup and strategy}\label{sec: setup}
\begin{figure}
	\centering
		
	\includegraphics[width = 12cm]{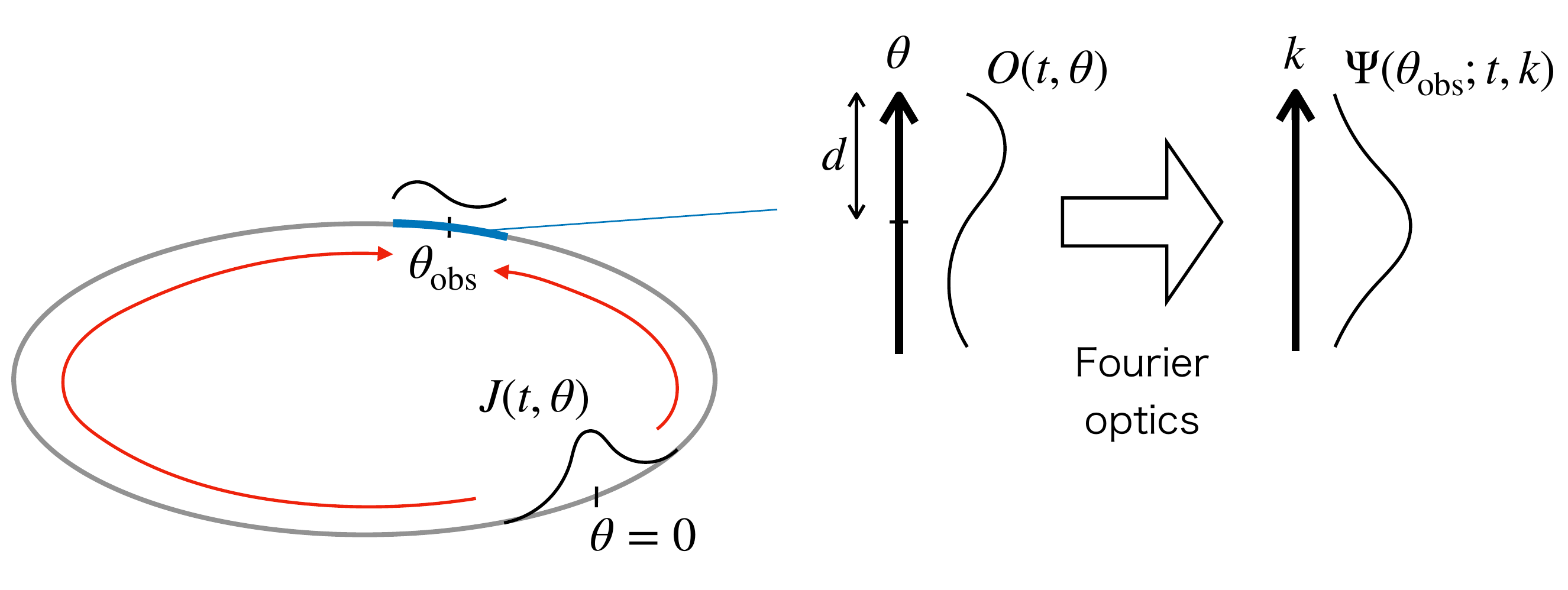}
	\caption{The setup and strategy.}
	\label{fig: setup}
\end{figure}
Our proposal to judge whether a ring-shaped material is an SEM is as follows (see Fig.\ref{fig: setup}).
Let $t$ denote the time coordinate and $\theta$ the coordinate along the ring with the periodic identification $\theta\sim \theta + a$, where $a$ is the circumference of the ring.
We put a gaussian source centered at $\theta = 0$,
\begin{align}
	J(t,\theta) = e^{-i\omega t} \frac{1}{\sqrt{8\pi \sigma^2}}\sum_n \exp\left(-\frac{(\theta-na)^2}{2\sigma^2 a^2} \right),
	\label{eq: source}
\end{align}
with the periodicity taken into account.\footnote{
The normalization factor is chosen to simplify later expressions. We have assumed the relativity and have set the material ``speed of light" to the unity. We have also assumed the isotropy of the material for simplicity, so that the speed of light does not depend on the direction of the propagation.}
The source is coupled linearly to the physical field $O(t,\theta)$ defined on the ring.\footnote{The corresponding experimental descriptions are found in Sec.~\ref{sec: experiment}.}
The effect of the source propagates over the ring, and we measure the observable $O(t,\theta)$, to which we mathematically apply the Fourier optics to obtain the image.
In the remaining of the section, we first introduce two models --- a model for an ordinary material and another model for a spacetime-emergent material --- for which in later sections we will check if the above strategy works, then illustrate the physical meaning of the Fourier optics.

\subsubsection*{Ordinary material}
In order to check if our strategy really works, we prepare two models without and with the emergent spacetime.
The first one is the free real scalar theory on the ring, with the source \eqref{eq: source} added.
The theory models scalar wave propagations in the simplest manner, and is known to roughly approximate Nambu-Goldstone modes accompanying spontaneous symmetry breaking when the matter is gapless.\footnote{
We would like to thank Youichi Yanase for a valuable advice on this point.
}
The action of the field $\phi(t,\theta)$, with the metric convention $\eta^{\mu\nu} = \mathrm{diag}(-1,1)$ and $\partial_\mu = (\partial_t,\partial_\theta)$, is given as
\begin{align}
	S_\mathrm{nongrav} = \int\d t\d \theta\,
	\l[-\frac{1}{2}\eta^{\mu\nu} \partial_\mu\phi(t,\theta)\partial_\nu\phi(t,\theta)-\frac{1}{2}m^2\phi(t,\theta)^2 + J(t,\theta)\phi(t,\theta) \r],
\end{align}
whose equation of motion (EOM) is derived by the $\phi$-variation:
\begin{align}
	(-\eta^{\mu\nu }\partial_\mu\partial_\nu  + m^2)\phi(t,\theta) = J(t,\theta).
	\label{eq: EOM of nongravity}
\end{align}

The model mimics the conductor/insulator transition characterized by the ratio of $m$ to $\omega$ (but note that the career is now the boson $\phi$).
As we will later compute in section \ref{sec: response}, the Green's function of \eqref{eq: EOM of nongravity} with frequency $\omega$ is given as
\begin{align}
	G(t,\theta) = \frac{1}{a}\sum_n \frac{e^{-i\omega t + ik_n\theta}}{k_n^2+m^2-\omega^2}\qquad 
	\l(k_n := \frac{2\pi n}{a}\r ).
	\label{eq: Green}
\end{align}
From this we can see that it has resonance modes around $k_n^2\sim \omega^2-m^2$ when $m<\omega$, while no resonance is found when $m>\omega$, which is the reason of the transition.

\subsubsection*{Spacetime-emergent material}
\begin{figure}
	\centering
		
	\includegraphics[width = 12cm]{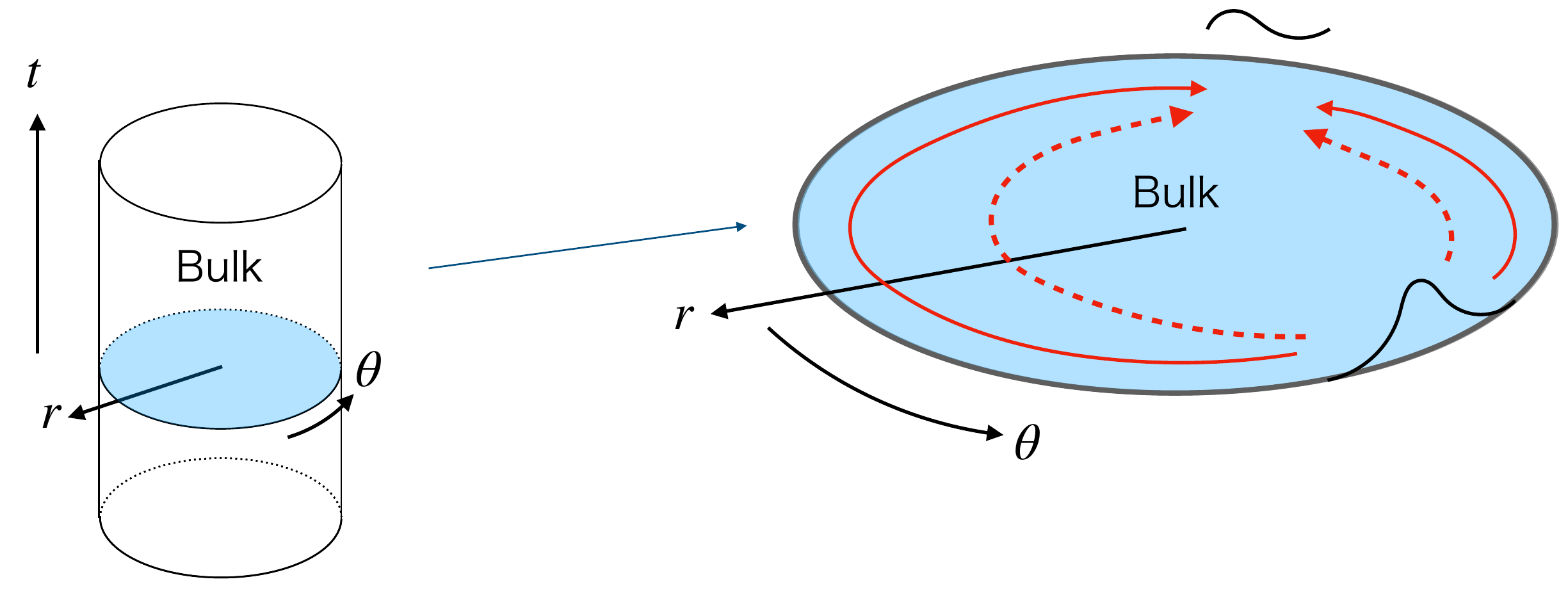}
	\caption{
	The emergent spacetime dual to the holographic CFT.
	In the holographic material, the response to the source behaves as if the source perturbation had propagated inside the virtual curved spacetime, the bulk.
	}
	\label{fig: holographic material}
\end{figure}

The second model is the simplest holographic CFT on the ring at nonzero temperature. 
Holographic CFT means that the CFT allows a bulk  gravitational description under the AdS/CFT correspondence.
Thus we describe this theory as an equivalent theory in the bulk.
We consider, as the dual bulk gravity theory, a massless scalar field $\Phi(x^0,x^1,x^2)$ on a 3-dimensional curved spacetime,
\begin{align}
	S_{\mathrm{grav}} = 
	-\frac{1}{2}\int\d x^3\sqrt{-g}\,g^{\mu\nu }
	\partial_\mu \Phi(x)\partial_\nu \Phi(x),
	\label{eq: action of gravity}
\end{align}
where  $x^\mu$ ($\mu=0,1,2$) is the spacetime coordinate, $\partial_\mu$ is $x^\mu$-derivative, $g_{\mu\nu }$ is the metric with norms of timelike vectors negative, $g^{\mu\nu }$ is its inverse and $g := \det g_{\mu\nu}$.
More specifically, the coordinates are $x^\mu = (t,\theta,r)$ with $r$ being the radial coordinate (see Fig.\ref{fig: holographic material}).
The radial coordinate is the emergent coordinate of the holographic spacetime. 

We adopt as $g_{\mu\nu}$ the following two spacetimes: the BTZ black hole spacetime and the AdS soliton spacetime.\footnote{In computing the response for $\Phi$ we keep the geometry $g_{\mu\nu}$ fixed (non-dynamical). This is called probe approximation, and it is justified since the source magnitude is supposed to be very small not to alter the temperature of the whole system, thus the back-reaction of the scalar field to the gravity field is ignored.}
The two spacetimes are solutions of the 3-dimensional Einstein equation with a negative cosmological constant, and as explained in section \ref{sec: introduction} there is a gravitational phase transition between them that are identified with the conductor/insulator transition in the material side.
The BTZ is favored when the temperature $T$ is higher than $1/a$, and the AdS soliton is favored when $T$ is lower than $1/a$.

In general, the explicit action of the holographic CFT is not available, while the duality tells that the two spacetimes correspond to some different states in a certain CFT. 
According to the AdS/CFT dictionary, the bulk scalar field $\Phi$ corresponds to a composite operator $\hat O(t,\theta)$ in the CFT, whose expectation value $O(t,\theta)$ is now affected by the source $J$ (given as \eqref{eq: source}) linearly coupled to the operator $\hat O(t,\theta)$.

The dictionary to derive the expectation value from the calculations in the gravity side is as follows.
The key tool connecting the bulk theory and the CFT is a dictionary called the GKPW relation \cite{Gubser:1998bc, Witten:1998qj}.
First, we solve the EOM from \eqref{eq: action of gravity},
\begin{align}
	\frac{1}{\sqrt{-g}}\partial_\mu (\sqrt{-g}g^{\mu\nu}\partial_\nu \Phi(x))=0,
	\label{eq: EOM of gravity}
\end{align}
to obtain the general solution.
Here the derivative operation acting on the bulk field $\Phi$ is the curved-spacetime version of the d'Alembertian.
Then, we find that the solution asymptotically behaves as
\begin{align}
	\Phi(x) \sim A(t,\theta) + \frac{B(t,\theta)}{r^2},
\end{align}
near the boundary.\footnote{
As we will see later in the detailed computation, this is not the precise expression in our setup though valid in most cases.
}
The dictionary \cite{Klebanov:1999tb} tells us to regard $A$ and $B$ as $J$ and $O$, respectively.

Thus, we adopt $A = J$ as one of the boundary conditions for our solution.
We also have to put another boundary condition inside the bulk to fix the remaining arbitrary constant; we postpone considering it to section \ref{sec: response}.
From those two conditions $O$ is uniquely determined as a function of $J$, which is the response function we want.\footnote{
Strictly speaking, the overall factor of $O$ is in general different from $B$. 
However, we are interested not in the overall factor, which will depend on the detail of the material, but only in the form of the function (the shape in its plot).
}

\subsubsection*{Imaging transform}
\begin{figure}
	\centering
		
	\includegraphics[width = 0.8\columnwidth]{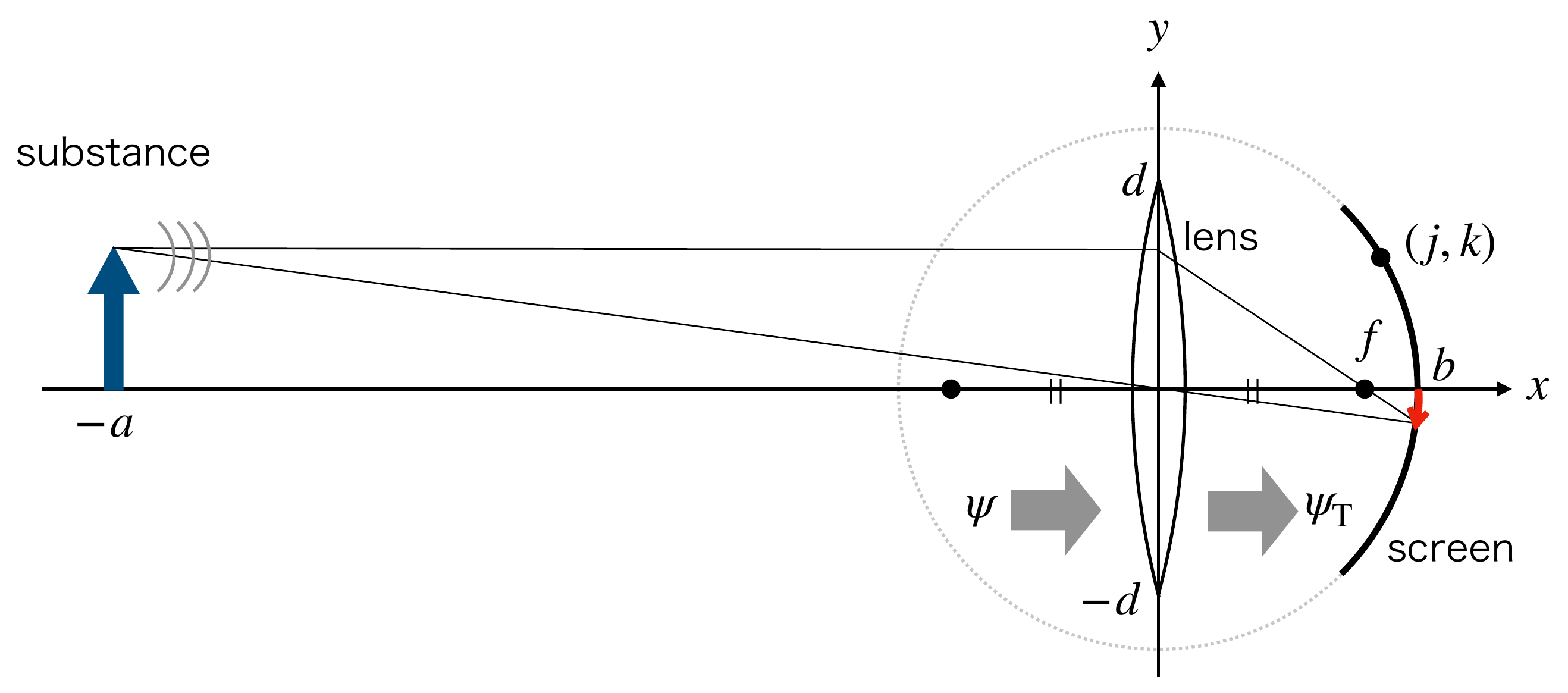}
	\caption{
	The imaging by lens. The lens is located at $x=0$, and the screen is at $x^2 + y^2 = b^2$, where $f$ is the focal length.
	$2d$ is the length of the lens, $a$ and $b$ is the distance from the lens to the substance and to the screen, respectively.
	The thin lines represent the light orbits in the geometrical optics approximation.
	}
	\label{fig: imaging}
\end{figure}
The method of Fourier optics realizes the visualization of the spacetime emergence. 
We apply the optical conversion formula just mathematically to the response function $O$, to visualize the bulk spacetime. Here we describe the mathematical formula, which is nothing but a Fourier transform.

To illustrate its physical meaning, let us first review how a substance is imaged by a convex lens and a screen as in Fig.\ref{fig: imaging}.
In the figure, spherical waves are emitted from points on the substance to reach the lens with the wave amplitude function being $\psi$, pass through the lens ($\psi_\mathrm{T}$), and then come to the screen ($\psi_\mathrm{S}$).
The lens gives a phase to the transmitted wave: $\psi_\mathrm{T}(y) = e^{-iy^2/2f}\psi(y)$.
The Huygens' theorem says that the image at the screen is composed of the superposition of wavelets coming from the lens.
Thus the wave at the screen is given as
\begin{align}
	\psi_{\mathrm{S}}(k) &\propto \int_{-d}^d\d y\,\psi_\mathrm{T}(y) \frac{e^{i\omega R(k,y)}}{R(k,y)},\qquad
	R(k,y) := \sqrt{j^2+(k-y)^2},
\end{align}
where $(j,k)$ is the cartesian coordinate on the screen, i.e., $j^2+k^2=b^2$.

We suppose $|y/b|\sim|k/b| \ll 1$, while $\omega y$ and $\omega k$ are kept $O(1)$.
Then $\psi_\mathrm{S}$ is approximated as
\begin{align}
	\psi_\mathrm{S}(k) \propto \int_{-d}^{d}\d y\,\psi(y)\exp\l[i\omega\left(\frac{y^2}{2a}-\frac{ky}{b} \right) \r ],
\end{align}
where the formula of lens, $1/f = 1/a + 1/b$, has been used.
Even when the screen is put at $b=f$, the image may still not be so blurred if we take $a/f\gg 1$.
In this case, we have
\begin{align}
	\psi_\mathrm{S}(k)  \propto \int_{-d}^{d}\d y\,\psi(y)e^{-i\omega ky/f}.
\end{align}
This is the mathematical imaging transform which is applied to the response function.
The form is nothing but a Fourier transform.

Let us go back to our theory setup and apply the above formula to the response $O$ as
\begin{align}
	\Psi(\theta_\mathrm{obs};t,k)=\int_{\theta_\mathrm{obs}-d}^{\theta_\mathrm{obs}+d}d\theta\, e^{-i\omega \theta k/f}O(t,\theta).
	\label{eq: imaging}
\end{align}
In this expression, $\Psi$ relates to $O$ just as $\psi$ does to $\psi_\mathrm{S}$; the convex lens is put at $\theta = \theta_\mathrm{obs}$ on the ring, with $y=\theta- \theta_\mathrm{obs}$ (as we have adopted the massless field in the bulk, the picture of the geometrical optics described above is still valid\footnote{
	We will discuss the case when the field is massive in section \ref{sec: discussion}.
}).
When the system has the emergent bulk spacetime, $O$ will behave as if a wave had propagated there.
In the picture of the geometrical optics, light rays are subject to the gravitational lens inside the bulk, which makes a virtual image for the observer at the boundary. 
The virtual image differs according to where the observer is located, and the lens functions as the observer's eyes.
Thus $\Psi$ contains geometric information about the bulk, which is why we claim that the imaging judges the spacetime emergence.\footnote{
Since the proper distance along the radial direction from a boundary point to a bulk point is always divergent, the condition $a/f\gg 1$ is automatically satisfied.
}
Actually, in higher-dimensional examples, the Einstein rings were visualized by using the same method \cite{Hashimoto:2018okj,Hashimoto:2019jmw, Zeng:2022woh, Kaku:2021xqp}. 
When the system does not have any dual gravity system, on the other hand, the function $\Psi$ does not allow any interpretation of the view of the bulk.


\section{Computation of response functions}\label{sec: response}
We compute the response function to the gaussian source \eqref{eq: source} for the two models introduced in section \ref{sec: setup}.

\subsection{Ordinary material}\label{subsec: response; ordinary}
Let us start with deriving the Green's function $G$ introduced in \eqref{eq: Green}, which satisfies
\begin{align}
	(-\eta^{\mu\nu }\partial_\mu\partial_\nu  + m^2)G(t,\theta) = \sum_{n=-\infty}^\infty \delta(\theta-n a),
\end{align}
where the r.h.s.\ is the identity in the space of functions with period $a$.
In experiments one expects dissipation and the mode with the source frequency $\omega$ will dominate the response as the stationary state (while the equation above does not respect the dissipation). 
Thus, we first put an ansatz $G(t,\theta) = e^{-i\omega t}g(\theta)$, to extract the dominant stationary part.
By Fourier-expanding $g(\theta)$ and the r.h.s.\ ($k_n$ is defined in \eqref{eq: Green}),
\begin{align}
	g(\theta) = \frac{1}{\sqrt{a}}\sum_n g_n e^{i k_n \theta },\qquad
	\sum_n \delta(\theta-na) = \frac{1}{a} \sum_n e^{i k_n\theta},\qquad 
\end{align}
we obtain
\begin{align}
	g_n = \frac{1}{\sqrt{a}}\frac{1}{k_n^2+m^2-\omega^2},\qquad
	\mathrm{i.e.,}\qquad
	g(\theta) = \frac{1}{a}\sum_n \frac{e^{ik_n\theta}}{k_n^2+m^2-\omega^2}.
\end{align}
From the Green's function, our solution of \eqref{eq: EOM of nongravity} is given as
\begin{align}
	\phi(t,\theta) = e^{-i\omega t}\int_{-\infty}^{\infty}\d \theta'\, g(\theta-\theta') J(\theta') = \frac{1}{2}\sum_n \frac{e^{-2\pi^2\sigma^2n^2}}{k_n^2+m^2-\omega^2}\, e^{-i\omega t + ik_n \theta}.
	\label{eq: phi}
\end{align}
This is the response function in our non-SEM model.

\subsection{Spacetime-emergent material}\label{subsec: response; SEM}
Here we compute the response function for our SEM model.
There is the phase transition of the background between the BTZ black hole (the high temperature phase, $T>1/a$) and the AdS soliton (the low temperature phase, $T<1/a$):
\begin{align}
	\mbox{BTZ:}\quad
	\d s_\mathrm{BTZ}^2 &= -\frac{r^2-r_h^2}{L^2}\d t^2 + \frac{L^2}{r^2-r_h^2}\d r^2 + \frac{r^2}{L^2}\d \theta^2,\qquad\qquad \label{eq: BTZ metric}\\
	\mbox{AdS soliton:}\quad
	\d s_\mathrm{sol}^2 &= -\frac{r^2}{L^2}\d t^2 + \frac{L^2}{r^2-r_s^2}\d r^2 + \frac{r^2-r_s^2}{L^2}\d \theta^2.\qquad\qquad \label{eq: soliton metric}
\end{align}
Here $L$ is the AdS radius, $r_h$ is the horizon radius, and the $r_s$ is the gap of the AdS soliton.
From the geometric regularity at $r=r_h$ and $r = r_s$ on the BTZ and the AdS soliton respectively, they are determined as
\begin{align}
	r_h = 2\pi L^2 T,\qquad\quad
	r_s = \frac{2\pi L^2}{a}.\label{eq: radii}
\end{align}
For $a = 2\pi L$ ($r_s = L$), the AdS soliton is identical to the pure AdS in the global patch.
Next, we compute the response function for each phase in order.

\subsubsection*{High temperature phase}
Let us first study the high temperature phase, the BTZ phase.
Since we are interested in the forced oscillation mode, we can expand the solution as
\begin{align}
	\Phi(x) = \sum_n \Phi_n(r) e^{-i\omega t+i k_n \theta}.
\end{align}
Putting
\begin{align}
	\Phi_n(r) = \xi^{-i\omega L^2/(2r_h)}f_n(\xi)\qquad\quad \l(\xi:=1-\frac{r_h^2}{r^2}\r ),
	\label{eq: def of f_n}
\end{align}
we obtain the simplified EOM,
\begin{align}
	\xi(1-\xi)f_n''(\xi) + \left(1-\frac{i\omega L^2}{r_h^2} \right)(1-\xi)f_n'(\xi) - \left(\frac{L^4n^2\pi^2}{a^2r_h^2}-\frac{L^4\omega^2}{4r_h^2} \right)f_n(\xi) = 0.
	\label{eq: simple EOM}
\end{align}
As this is the hypergeometric equation, the general solution around $\xi = 0\,(r=r_h)$ is written as
\begin{align}
	f_n(\xi) = C_n^1 F(\alpha_n,\beta_n,\gamma_n;\xi) + C_n^2\xi^{1-\gamma_n}F(\alpha_n-\gamma_n+1,\beta_n-\gamma_n,2-\gamma_n;\xi),
	\label{eq: BTZ f_n}
\end{align}
where $F$ is the hypergeometric function, and $\alpha_n,\beta_n$ and $\gamma_n$ are defined as
\begin{align}
	\alpha_n := -i\left(\frac{\omega L^2}{2r_h}+\frac{n\pi L^2}{ar_h} \right),\quad
	\beta_n := -i\left(\frac{\omega L^2}{2r_h}-\frac{n\pi L^2}{ar_h} \right),\quad
	\gamma_n := 1+\alpha_n+\beta_n.
	\label{eq: alpha beta}
\end{align}

We determine the coefficients $C_n^1$ and $C_n^2$ by imposing the following two boundary conditions on the solution.
First, we require the in-going boundary condition at the black hole horizon. 
In $\Phi_n$ given by \eqref{eq: def of f_n} and \eqref{eq: BTZ f_n}, we have the near-horizon expression
\begin{align}
		e^{-i\omega t}\xi^{\mp i\omega L^2/(2r_h)} = \exp\left(-i\omega t\mp \frac{i\omega L^2}{2r_h}\ln\xi \right)
	\simeq \exp\left[-i\omega\l( t\pm \frac{L^2}{2r_h}\ln\frac{r-r_h}{r_h} \r)\right],
\end{align}
which leads us to impose $C_n^2 =0$, i.e.,
\begin{align}
	\Phi(x) = \sum_{n=-\infty}^{\infty}C_n^1e^{-i\omega t+ik_n\theta} \xi^{-i\omega L^2/(2r_h)}F(\alpha_n,\beta_n,\gamma_n;\xi).
	\label{eq: BTZ Phi with C_1}
\end{align}

The other condition is that the non-normalizable mode in the bulk corresponds to the source in the boundary theory, according to the AdS/CFT dictionary, as was explained in section 2.
The asymptotic form of \eqref{eq: BTZ Phi with C_1} near the boundary of the bulk $\xi \sim 1$ is, at the leading order,
\begin{align}
	 \Phi(x) &\sim \sum_{n=-\infty}^{\infty}C_n^1e^{-i\omega t+ik_n\theta}.
\end{align}
By equating this to $J$ of \eqref{eq: source} (with $e^{-i\omega t}$ removed), we have
\begin{align}
	\sum_n C_n^1 e^{ik_n\theta} 
	= \frac{1}{\sqrt{8\pi \sigma^2}}\sum_n \exp\left(-\frac{(x-na)^2}{2\sigma^2 a^2} \right).
\end{align}
Then the Fourier coefficient $C_n^1$ is determined as
\begin{align}
 	C_n^1 \simeq \frac{1}{2}e^{-2\pi^2\sigma^2n^2}\qquad (\sigma\ll 1),
 	\label{eq: BTZ C_1}
\end{align}
where we have replaced the integral domain $[-a/2,a/2]$ with $[-\infty,\infty]$ by supposing that the source is local.

Next, we read the response from the subleading term in the near-boundary expansion.
By solving the asymptotic form of the EOM \eqref{eq: EOM of gravity}, one finds two modes $r^{0}$ and $r^{-2}$, as mentioned in section \ref{sec: setup}.
The former is non-normalizable and corresponds to the source, while the latter is normalizable and corresponds to the response.
Here we need to take care of a subtlety in the expansion: in fact, the correct expansion of \eqref{eq: BTZ Phi with C_1} with \eqref{eq: BTZ C_1} is, to $O(r^{-2})$, given as
\begin{align}
  \Phi(x) &= \frac{1}{2}\sum_n e^{-i\omega t+2ik_n\theta-2\pi^2\sigma^2n^2}\xi^{-i\omega L^2/(2r_h)}F(\alpha_n,\beta_n,\gamma_n;\xi)\nonumber \\
  &\sim \frac{1}{2}\sum_{n=-\infty}^{\infty}e^{-i\omega t+ik_n\theta-2\pi^2\sigma^2n^2}\nonumber\\
  &\quad\times \left[1 + \frac{i\omega L^2r_h}{2}\epsilon + r_h^2\alpha_n\beta_n \epsilon\ln(r_h^2\epsilon)+\alpha_n\beta_n\left(H_{\alpha_n}+H_{\beta_n}-1 \right)r_h^2 \epsilon+ O(\epsilon^2\ln\epsilon)\right],
  \label{eq: BTZ expansion}	
\end{align}
where we have put $\epsilon:=1/r^2$ and $H_p$ is the analytically continued harmonic number.\footnote{
$H_p$ is related to the digamma function $\psi(p)$ as $H_p = \gamma + \psi(p+1)$, where $\gamma$ is the Euler's constant.
}
Then we find that the story seems more complex than we expected.

The problem is caused by that the non-normalizable mode has $O(r^{-2}\ln r)$ and $O(r^{-2})$ terms as its next orders in solving the asymptotic EOM and they are more (as) dominant than (as) the normalizable mode.
If we solve the asymptotic EOM to $O(r^{-2})$, the linear combination of the two mode reads
\begin{align}
	\sum_n e^{-i\omega t+ik_n\theta}\l[D_n^1(1+\alpha_n\beta_n r_h^2\epsilon\ln(r_h^2\epsilon)-(1+\alpha_n\beta_n)r_h^2\epsilon) + D_n^2\epsilon + O(\epsilon^2\ln\epsilon)\r].
\end{align}
We can determine $D_n^{1,2}$ by comparing this with \eqref{eq: BTZ expansion}.
The response can be read from the normalizable mode, the sum of the $D_n^2$ terms.
Since we find that $D_n^1$ is identical to $C_n^1$ by construction, we obtain
\begin{align}
	D_n^2 = 	\frac{r_h^2}{2}e^{-2\pi^2\sigma^2n^2}\l[1 + \frac{i\omega L^2}{2r_h}+\alpha_n\beta_n(H_{\alpha_n}+H_{\beta_n})\r],
\end{align}
and hence we conclude that the response function in the BTZ phase, the high temperature phase ($T>1/a$), is
\begin{align}
  O_{\mathrm{BTZ}}(t,\theta) = \frac{r_h^2}{2}\sum_n e^{-i\omega t+ik_n\theta-2\pi^2\sigma^2n^2}\l[1 + \frac{i\omega L^2}{2r_h}+\alpha_n\beta_n(H_{\alpha_n}+H_{\beta_n})\r].
  \label{eq: BTZ response}
\end{align}

\subsubsection*{Low temperature phase}
The computation in the low temperature phase, the AdS soliton phase, is similar to the previous case.
We obtain the same solution as \eqref{eq: BTZ f_n}, but with
\begin{align}
	\Phi_n(r) = \xi^{\pi |n|L^2/(a r_s)}f_n(\xi)\qquad\quad \left(\xi = 1-\frac{r^2_s}{r^2} \right),
\end{align}
and 
\begin{align}
	\alpha_n := \frac{\pi |n|L^2}{ar_s}-\frac{\omega L^2}{2r_s},\qquad
	\beta_n := \frac{\pi|n| L^2}{a r_s}+\frac{\omega L^2}{2r_s},\qquad
	\gamma_n := 1+\alpha_n+\beta_n,
	\label{eq: sol alpha beta}
\end{align}
instead of \eqref{eq: def of f_n} and \eqref{eq: alpha beta}, respectively.

As the boundary condition inside the bulk, in the previous AdS black hole case we considered the in-going boundary condition at the black hole horizon.
Now, in the present AdS soliton geometry, the spacetime does not contain any black hole and is regular at $r=r_s$, so we require that $\Phi$ is not divergent at $r=r_s$.
From this, we again get $C^2_n=0$.
The response can be computed by following the same procedure as before.
The response function in the AdS soliton phase, the low temperature phase ($T<1/a$), is
\begin{align}
	O_{\mathrm{sol}}(t,\theta) = \frac{r_s^2}{2}\sum_n e^{-i\omega t+ik_n\theta-2\pi^2\sigma^2n^2}\l[1 - \frac{\pi|n| L^2}{2ar_s}+\alpha_n\beta_n(H_{\alpha_n}+H_{\beta_n})\r].
	\label{eq: soliton response}
\end{align}
Note that this $O_\mathrm{sol}$ does not have the $T$-dependence.


\section{Imaging response functions}\label{sec: imaging}

\begin{figure}[p]
\centering
	
	\begin{tabular}{cc}
		\begin{minipage}{0.45\columnwidth}
			\includegraphics[height = 5cm]{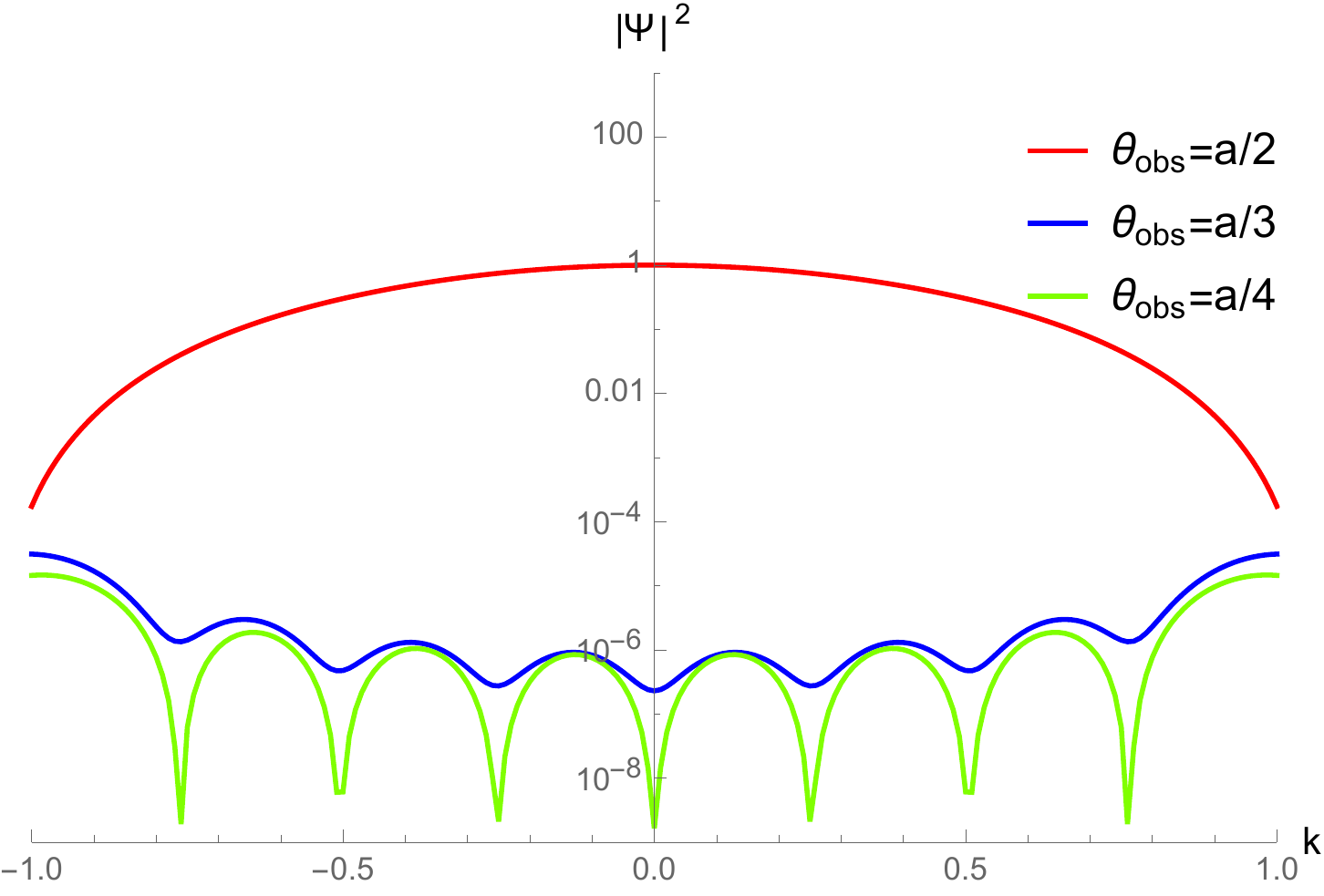}
		\end{minipage}
        &
		\begin{minipage}{0.45\columnwidth}
			\includegraphics[height = 5cm]{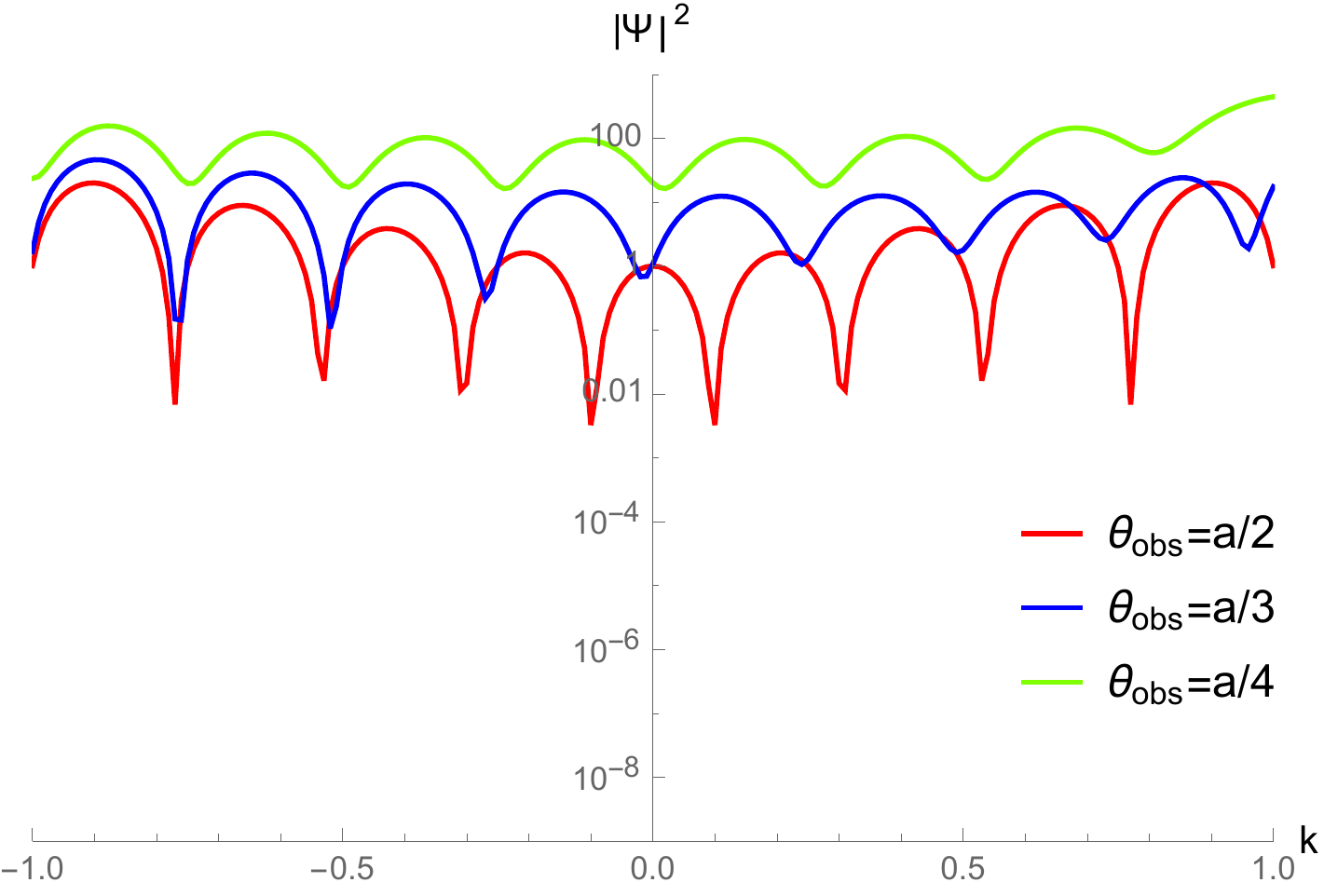}
		\end{minipage}\\[6pt]
        {\footnotesize SEM (low $T$, AdS soliton)} &{\footnotesize SEM (high $T$, BTZ)}

    \end{tabular}\\[12pt]
    
    \begin{tabular}{cc}
        \begin{minipage}{0.45\columnwidth}
            \includegraphics[height = 5cm]{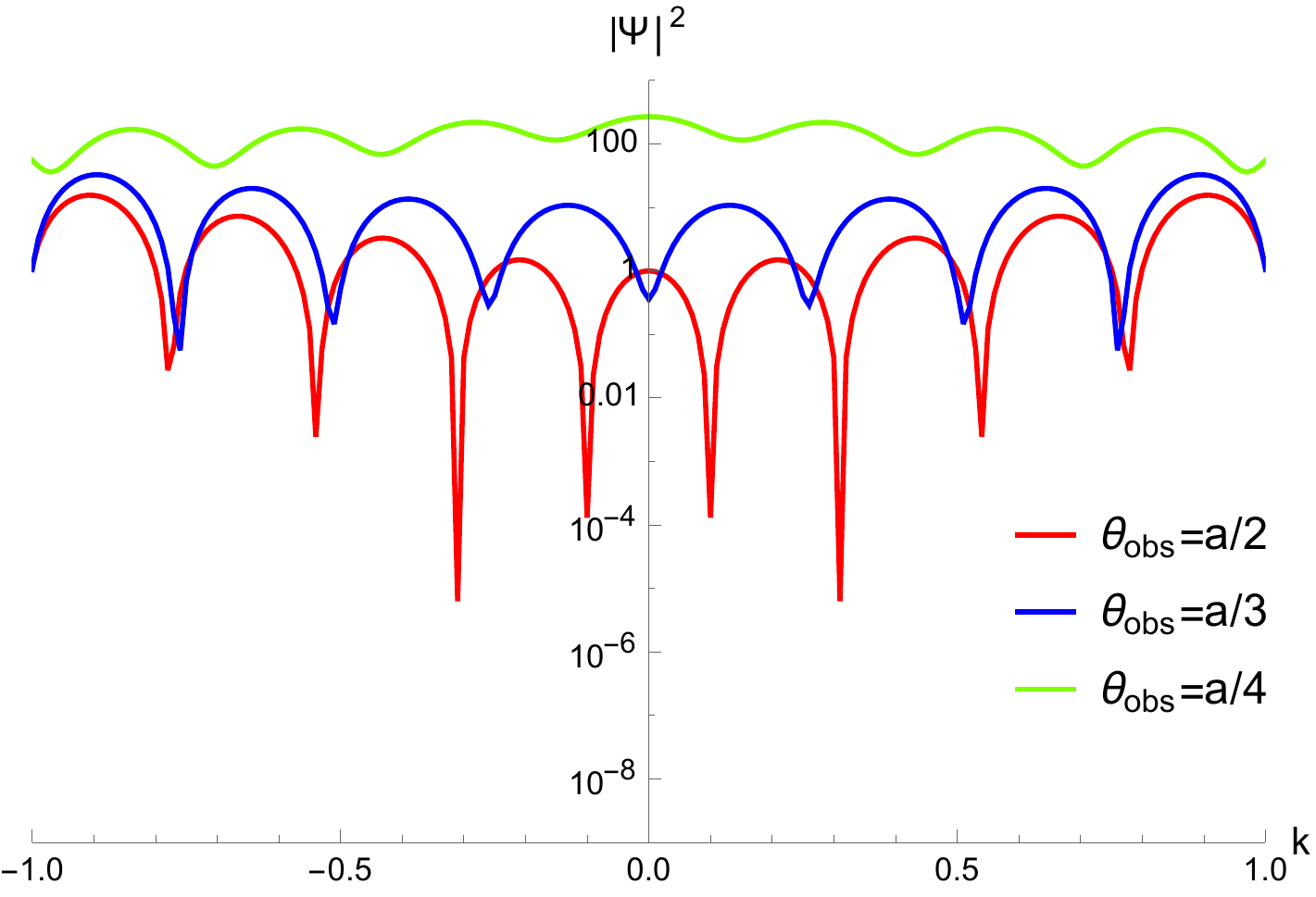}
		\end{minipage}
        &
        \begin{minipage}{0.45\columnwidth}
            \includegraphics[height = 5cm]{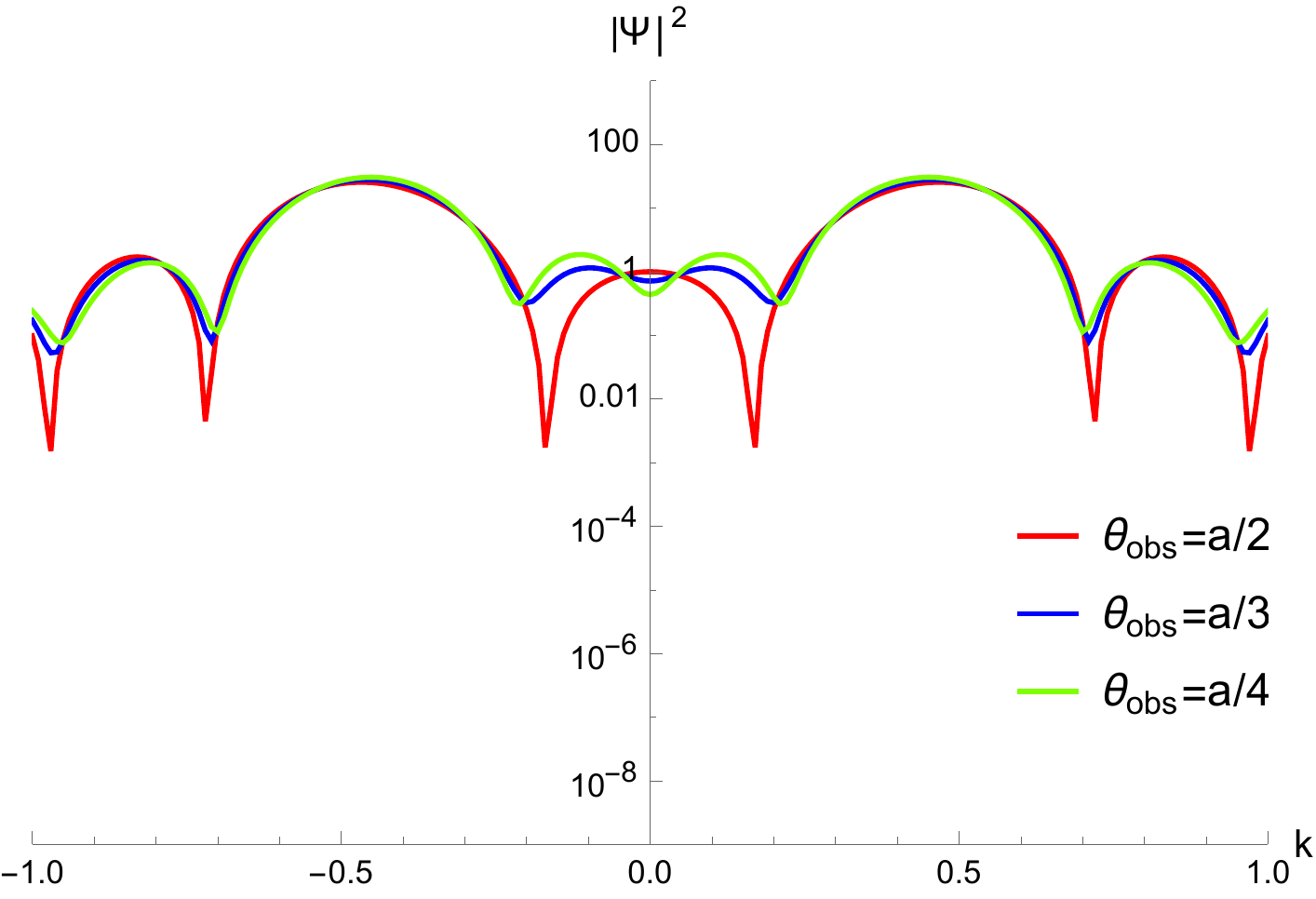}
		\end{minipage}\\[6pt]
        {\footnotesize non-SEM (low $T$, $m=55\pi/a>\omega$)} & {\footnotesize non-SEM (high $T$, $m=45\pi/a<\omega$)}
	\end{tabular}\vspace{36pt}
	\caption{
	The plots of the images of the response functions.
    The upper row is for SEM and the lower for non-SEM.
    On the other hand, the left column is for low temperature phase and the right for high (though the non-SEM model does not depend on temperature, the transition due to the ratio $m/\omega$ mimics the conductor/insulator transition, which is usually determined by temperature).
    Three plots on each panel are normalized so that the plot function for $\theta_\mathrm{obs}=a/2$ at $\theta=0$ is set to be 1.
	(Parameters, except $\sigma = 10^{-2}$ (the width of $J$ is $a\sigma$), are chosen so that $a$ determines their dimension: $\omega = 101\pi/(2a)$, $d = a/(4\pi)$, and the others are shown above.
	Because of this choice, $a$ contributes to $\Psi$ only as its overall factor, thus we can set $a$ arbitrarily.)
	}
	\label{fig: transformed image}
\end{figure}

\begin{figure}
    \centering
    	
    \includegraphics[width = 0.9\columnwidth]{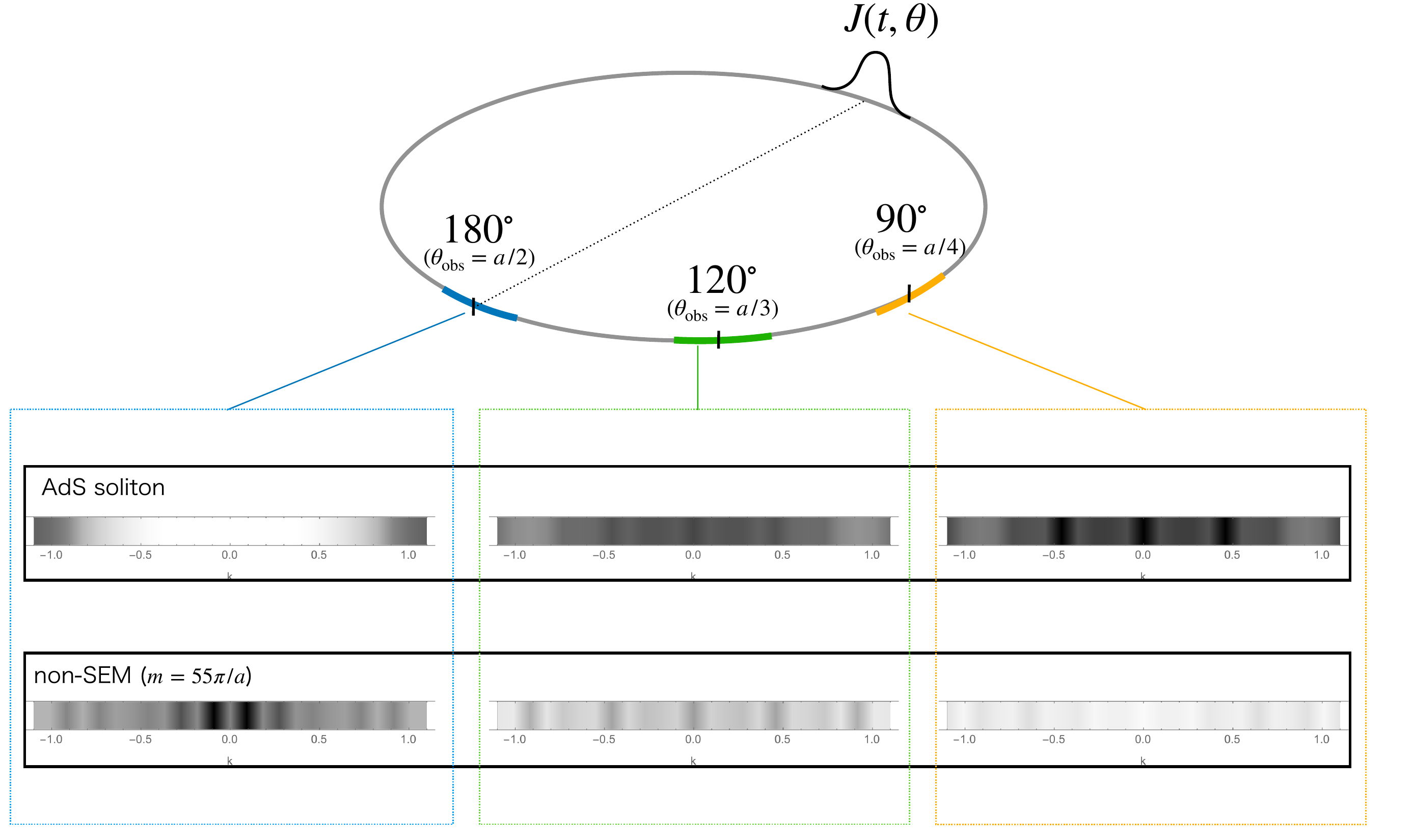}
    \caption{The density plots of the images of the response functions at $\theta_{\rm obs}=a/2, a/3,$ and $a/4$. The upper row is for the SEM model at the low temperature phase (AdS soliton), and we plot a function $(4+\log_{10}|\Psi|^2)/7$ which roughly ranges from 0 to 1 for the computed image values.  The lower row is for the non-SEM model (with the scalar mass $n=55\pi/a$), where we plot $(12+\log_{10}|\Psi|^2)/7$ which again roughly ranges from 0 to 1. }
    \label{fig:lightdark plot}
\end{figure}

We apply the imaging transform \eqref{eq: imaging} to $\phi$ in \eqref{eq: phi}, $O_\mathrm{BTZ}$ in \eqref{eq: BTZ response} and $O_\mathrm{sol}$ in \eqref{eq: soliton response}.
Since they depend on $k$ only through $k/f$, we can set $f=1$.
The result is as follows:
\begin{align}
	\Psi_\phi(\theta_\mathrm{obs};t,\theta) &= \sum_n\frac{e^{-2\pi^2\sigma^2n^2}}{k_n^2+m^2-\omega^2}e^{i\theta_\mathrm{obs}(k_n-\omega k)}\frac{\sin(d(k_n-\omega k))}{k_n-\omega k},\\
	\Psi_\mathrm{BTZ}(\theta_\mathrm{obs};t,\theta) &= 
	r_h^2\sum_n e^{-2\pi^2\sigma^2 n^2}\l[1 + \frac{i\omega L^2}{2r_h}+\alpha_n\beta_n(H_{\alpha_n}+H_{\beta_n})\r]_\mathrm{Eq.\eqref{eq: alpha beta}} \nonumber\\
	&\qquad\qquad \times e^{i\theta_\mathrm{obs}(k_n-\omega k)}\frac{\sin(d(k_n-\omega k))}{k_n-\omega k},\\
	\Psi_\mathrm{sol}(\theta_\mathrm{obs};t,\theta) &= 
	r_s^2\sum_n e^{-2\pi^2\sigma^2 n^2}\l[1 - \frac{\pi|n| L^2}{2ar_s}+\alpha_n\beta_n(H_{\alpha_n}+H_{\beta_n})\r]_\mathrm{Eq.\eqref{eq: sol alpha beta}}\nonumber\\
	&\qquad\qquad \times e^{i\theta_\mathrm{obs}(k_n-\omega k)}\frac{\sin(d(k_n-\omega k))}{k_n-\omega k}.
\end{align}
Recalling \eqref{eq: radii}, we see that the AdS radius $L$ contributes only as an overall factor, so we also set $L=1$.

Fig.\ref{fig: transformed image} shows the plot of each $|\Psi(\theta_\mathrm{obs};t,\theta)|^2$, which is independent of $t$ any more.
We are able to identify the image of the low temperature phase of SEM (the AdS soliton phase) from the others, since its $\theta_\mathrm{obs}$-dependence of the response image is the most dramatic, while the others do not have such a strong dependence on $\theta_\mathrm{obs}$.
For the AdS soliton, we catch a strong signal at $\theta_\mathrm{obs} = a/2$, while it cannot be seen at
non-antipodal points ($\theta_\mathrm{obs}\neq a/2$), embodying the emergence of the bulk spacetime by ``seeing" the source through the bulk.\footnote{The peculiarity of the AdS soliton phase with the strong signal can be seen also in the response function itself. See App.~\ref{app:A} for the explicit plot of the response functions.} 
In the high temperature phase of the SEM (the BTZ phase), the image is hardly different from that of the non-SEM with large masses,\footnote{Note that the magnitude of the image function for the non-SEM model shows the ordinary expected transition from the conducting phase to the insulator phase when the temperature is lowered (or equivalently the mass $m$ is increased against $\omega$).} and hence we need to explore the low temperature phase in order to judge if a spacetime is emergent.

In Fig.\ref{fig:lightdark plot}, we show the density plots of Fig.\ref{fig: transformed image}, for the SEM model at low temperature (the AdS soliton case) and for the non-SEM model with $m=55\pi/a$. 
For the AdS soliton, the image is bright only at the antipodal point, while for the non-SEM model the images get brighter when the observation point $\theta_{\rm obs}$ approaches the source.

\begin{figure}
	\centering
		
	\begin{minipage}[b]{0.4\columnwidth}
		\centering
		\includegraphics[width = 6cm]{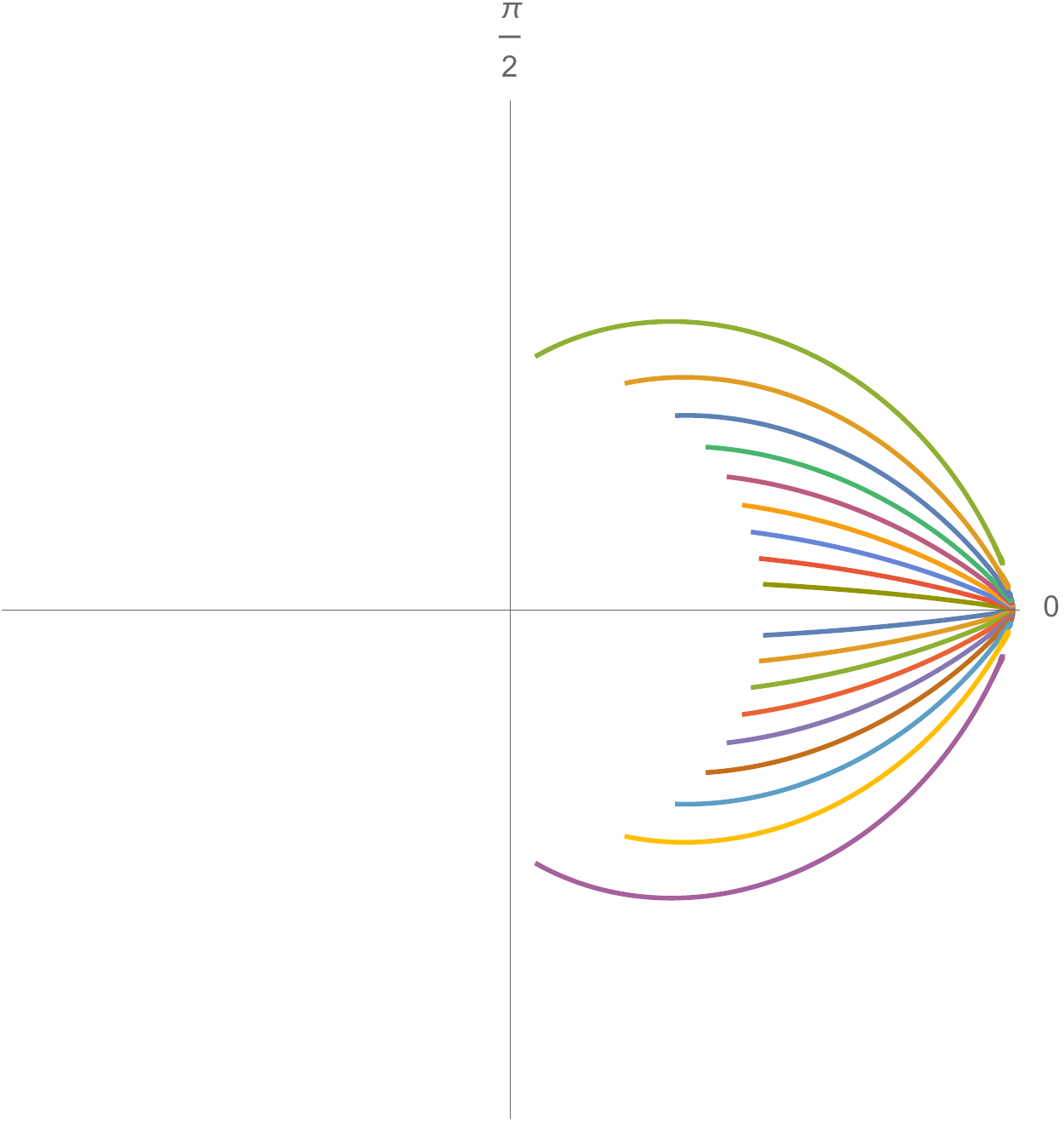}
		\caption*{BTZ}
	\end{minipage}
	\centering
	\begin{minipage}[b]{0.4\columnwidth}
		\centering
		\includegraphics[width = 6cm]{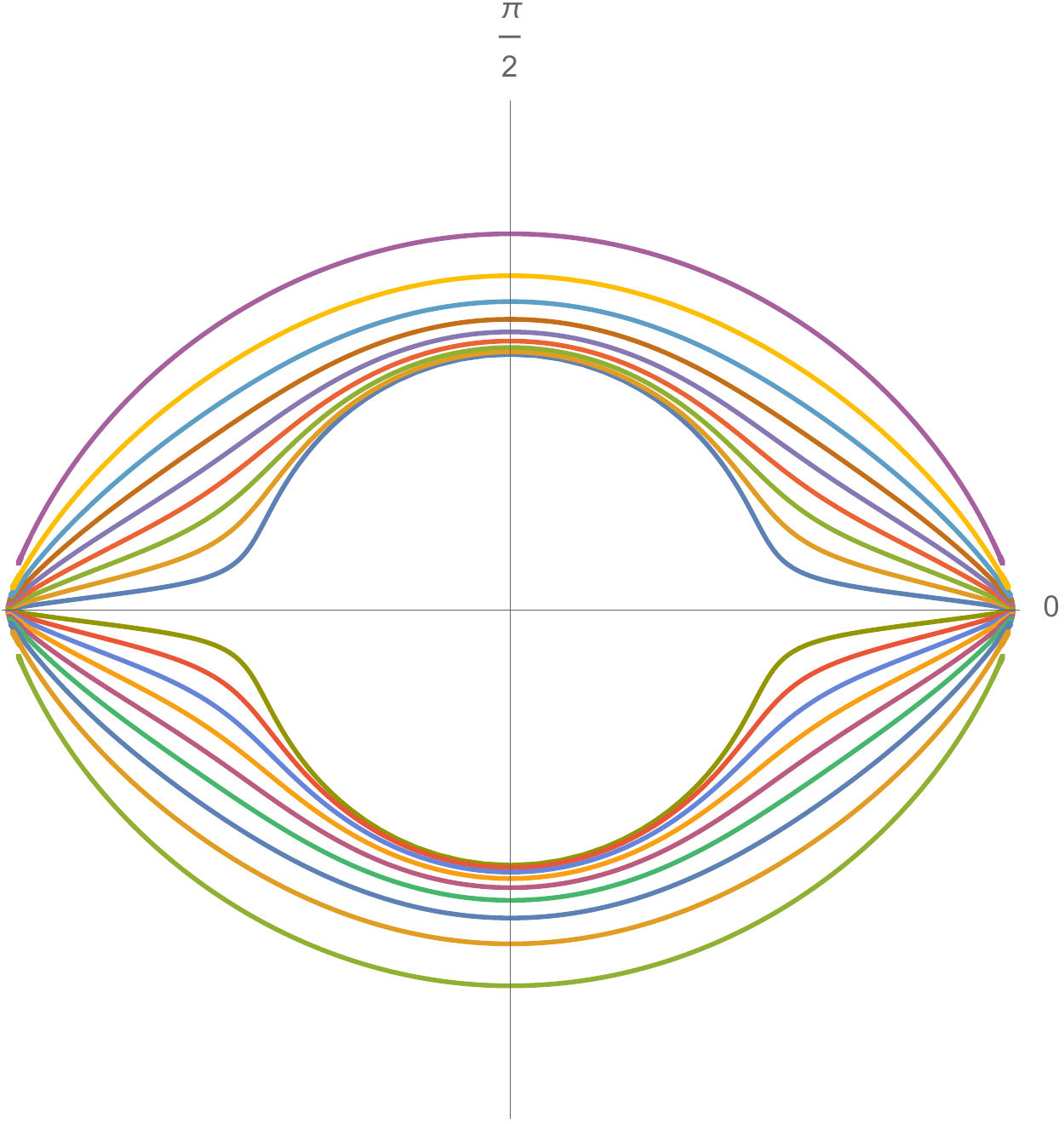}
		\caption*{AdS soliton}
	\end{minipage}
	\caption{
	Null geodesics shot from a boundary point, projected onto the $(r,\theta)$-plane.
	The two axis labels indicate the value of $2\pi \theta/a$ in those directions.
	In these diagrams the radial direction is scaled to show the spatial boundary $r=\infty$ at a finite range, by a new radial coordinate $(2/\pi)\tan^{-1} r$.
	The null geodesics in the BTZ spacetime reach the black hole horizon at $\lambda = -r_h/(1-l^2)$.
	}
	\label{fig: null geodesics}
\end{figure}

The behavior of the response image of the SEM model allows in fact the geometric interpretation of the emergent spacetime. Let us consider the geometrical optics approximation (the geodesic analysis).
From \eqref{eq: BTZ metric} and \eqref{eq: soliton metric}, null geodesics shot from the boundary point are respectively given as 
\begin{align}
	\mbox{BTZ:}
	\quad t(\lambda) &= \frac{1}{r_h}\coth^{-1}\left(\frac{(1-l^2)\lambda}{r_h} \right),\nonumber\\
	\theta(\lambda) &= \frac{1}{r_h}\coth^{-1}\left(\frac{(1-l^2)\lambda}{lr_h} \right),\quad
	r(\lambda) = \sqrt{(1-l^2)\lambda^2 - \frac{r_h^2l^2}{1-l^2}}\\
	\mbox{AdS soliton:}
	\quad t(\lambda)&=\frac{1}{r_s}\tan^{-1}\left(\frac{(1-l^2)\lambda}{r_s} \right)+\frac{\pi}{2r_s}, \nonumber\\
	\theta(\lambda) &= \frac{1}{r_s}\tan^{-1}\left(\frac{(1-l^2)\lambda}{lr_s} \right)+\frac{\pi}{2r_s},\quad
		r(\lambda) = \sqrt{(1-l^2)\lambda^2 + \frac{r_s^2}{1-l^2}},
\end{align}
where $l\in (-1,1)$ is the parameter classifying null geodesics, and $\lambda$ is the worldline coordinate which, when eliminated through $(t(\lambda), \theta(\lambda), r(\lambda))$, provides the bulk geodesic $\theta(r)$.
Each geodesic starts at $(t,\theta,r) = (0,0,\infty)$, where $\lambda = -\infty$.
The two geodesic families are shown in Fig.\ref{fig: null geodesics}.
In the BTZ spacetime, all light rays go down into the horizon, and never come back to the boundary.
In the AdS soliton spacetime, on the other hand, all light rays are accumulated to reach the antipodal point, vertically to the boundary line due to the gravitational lens; in the AdS soliton case, $\d\theta/\d r\to 0$ as $\lambda\to \pm \infty$. The image view consists of a bright spot of light coming from the front.  
Thus in this sense, we can say that the imaging transform visualizes the spacetime emergence.\footnote{In experiments with real materials, the emergent geometry may not exactly be the AdS soliton geometry; it could be some deformed geometry. We will discuss this issue in section \ref{sec: discussion}.}

\section{Material parameters for search experiment}
\label{sec: experiment}

As we have emphasized in the introduction, ring-shaped materials are suitable in searching spacetime-emergent materials.
Here in this section we look at the parameters of the theory toward the experimental realization. The parameters are the circumference
of the ring $a$, the angular frequency of the source $\omega$, and the temperature $T$. The realization 
highly depends on the actual values of these parameters, and in particular depends on what kind of
waves we consider on the materials. Below we explicitly propose TlCuCl$_3$ for the material, and calculate the values of the parameters for possible experiments.

Among the quantum critical points (QCPs) observed and proposed in literature, one of the suitable 
QCPs for our purpose is that related to spin waves (magnons).\footnote{
The rigidity of our result against the change of the career's spin is discussed in section \ref{sec: discussion}.
}
In this case, the observable operator ${\cal O}$ is the spin density operator, and $J$, the source linearly coupled to ${\cal O}$, could be the electromagnetic source. 
In the actual experiment, the Fourier optics imaging can be performed numerically from data sets of ${\cal O}(\theta)$.

There are various reasons for choosing the spin wave, as listed below. 
\begin{itemize}
\item
The waves conduct only on the ring and do not propagate in the physical inside of the ring.
If we have used
electromagnetic waves instead of the spin waves, they would propagate inside the ring and the holographic emergence cannot be checked. 
\item
We can get closer to the QCP.  If we have instead used a high $T_{\rm c}$ superconductor, commonly studied in the field of the AdS/CMP, 
the QCP would have been hidden inside the superconducting dome in the parameter space.
\item
We can use electrical insulators. If we have used conducting materials, the electromagnetic waves used for controlling
the source would also affect the conduction itself.
\item
The coherence length for spin waves can be long enough so that the continuum approximation of
our theory can be justified.
\end{itemize}
For these reasons, we come to consider thallium copper chloride, TlCuCl$_3$.
It is one of the popular materials that exhibit the quantum phase transition, which can be controlled not by the material compositions but by the external magnetic field or the pressure \cite{matsumoto2002magnon,Matsumoto2004FieldAP} (see also a review article \cite{sachdev2011quantum} which mentions the material and the application to the holographic principle).
Below, we use material parameters of the TlCuCl$_3$ to check whether the consistency conditions of our theory setup is satisfied or not, to see the experimental realization.

Our theory, in particular the analyses of the AdS soliton geometry, needs the following two conditions. First, the imaging condition 
\begin{align}
    \omega \gg \frac{2\pi v}{a}
    \label{eq:imaging_cond}
\end{align}
is required so that the imaging resolution is high. This condition means that the wave length on the
ring material is small compared to the size of the lens region. Here we have restored $v$ which 
is the speed of the wave propagation on the ring.\footnote{In the theoretical analyses in previous sections we have put $v=1$ as it corresponds to the speed of light, following high energy theory notations.
For numerical simulations, we took $\omega = 101 \pi v/(2a) \gg 2\pi v/a$.}
Second, we require the low-temperature condition 
\begin{align}
    \frac{v \hbar}{a} > k_{\rm B}T, 
        \label{eq:soliton_cond}
\end{align}
since we are going to probe the AdS soliton geometry which is realized at the low temperature.
The threshold is located at the phase transition temperature separating the soliton phase 
from the BTZ black hole phase, $T=1/a$ in the unit of $v=\hbar=k_{\rm B}=1$.
These two conditions \eqref{eq:imaging_cond} and \eqref{eq:soliton_cond} are the necessary conditions
for the theory analyses to be valid.

Let us substitute material values for these conditions. With the velocity of the magnon on the material TlCuCl${}_3$ estimated roughly as 
$v \simeq 2 \times 10^3$ m/s,\footnote{The magnon speed $v$ can be estimated as follows. According to
the measured dispersion relation described in Fig.~2 of \cite{cavadini2001magnetic},
at the near-gapless
point C $=(0,0,1)$ in the reciprocal space, the slope of the dispersion is evaluated as $2.1$ meV$\mathrm{\mathring{A}}$ for the $(1,0,-2)$ direction, $12$ meV$\mathrm{\mathring{A}}$ along the $b^*$ axis, and $21$ meV$\mathrm{\mathring{A}}$  along the $c^*$ axis, respectively. These slopes correspond to the propagation speed of $3.1 \times 10^2$ m/s, $1.8 \times 10^3$ m/s, and $3.3 \times 10^3$ m/s respectively. Thus if the ring is made from a flake whose in-plane directions are away from the $(1,0,-2)$ direction (i.e. the direction with the slowest magnon velocity), the magnon dispersion can be reasonably isotropic in the ring plane of the material.
This can be done, for example, by cleaving the material along the $(1,0,-2)$ plane, which is one of the cleavage plane. Then, the propagation speed in the ring is rather isotropic with the value around $v \sim 2 \times 10^3$ m/s. We use this value for the evaluation of the experiment parameters.
} we consider the following three cases for the temperature value for the SEM search experiment.
\begin{itemize}
\item 
    $T\simeq 1.5$ K (which is the value considered in \cite{matsumoto2002magnon,Matsumoto2004FieldAP}). From \eqref{eq:soliton_cond} we get $a\lesssim 10$ nm, which is so small that the lattice effect of the material cannot be ignored, since the lattice constant\footnote{According to \cite{tanaka2001observation}, the lattice parameters of TlCuCl$_3$ at room temperature are $b=14.1440 \mathrm{\mathring{A}}$ and $c=8.8904 \mathrm{\mathring{A}}$.} of TlCuCl$_3$ is about $1$ nm.
    For this maximum value $a \sim 10$ nm, the condition
    \eqref{eq:imaging_cond}  gives the frequency $\nu \equiv \omega/2\pi \gg 10^2$ GHz, and a special experimental care may be necessary to prepare the source.
\item 
    $T\simeq 10$ K. From \eqref{eq:soliton_cond} we get $a\lesssim 1$ nm which is too small for the continuum theory to be valid.
\item 
    $T\simeq 0.1$ K. From \eqref{eq:soliton_cond} we get $a \lesssim 10^2$ nm.
        This is much larger than the lattice constant of the material, thus our continuum limit is justified. The condition
    \eqref{eq:imaging_cond}  gives the frequency $\nu \equiv \omega/2\pi \gg 10$ GHz. This value is accessible for the source to be prepared by electromagnetic method.
\end{itemize}
A general relation found here is summarized as follows. For temperature $T\sim 10^n$ K, the ring circumference needs 
to satisfy $a \lesssim 10^{1-n}$ nm, and the frequency $\nu$ of the source needs to satisfy $\nu \gg 10^{n+2}$ GHz.
Therefore the material parameters for our theory analyses to be valid allow a small window: $n\lesssim -1$.

These rough calculations show that
the theory conditions \eqref{eq:imaging_cond} and \eqref{eq:soliton_cond} 
are satisfied at low temperature experiments, providing a realistic possibility of the experimental search for the
spacetime-emergent materials.

\section{Summary and discussions}
\label{sec: discussion}

In this work, we have put a theoretical basis for future experiments for finding spacetime-emergent materials (SEMs).
When a local source is put at a point on the ring-shaped material, the response function behaves differently according to whether or not the material of interest is an SEM. The difference is manifested by the visualization using the imaging transform of the response.
The most distinguishable feature is that for the case of the SEM, the image shows a strong signal at the antipodal point of the ring when the material is cooled enough.
This feature in fact allows the geometric interpretation of the emergent spacetime, since in our calculated examples the image coincides with the expectation from the geometrical optics approximation in the bulk (geodesic analysis). Our method provides the image as if the observer looks into the holographically emergent curved spacetime.
Thus, we claim that this imaging method with the ring-shaped material prepared can verify the spacetime emergence.

We have also discussed possible directions for realizing our strategy in laboratories, raising a candidate material for the SEM. 
The material TlCuCl$_3$, which has a quantum critical point at zero temperature, allows enough amount of material information including the magnon speed and lattice structure. It has enabled us to estimate experimental parameters for the search of SEMs, for example at the temperature $T=$ 0.1 K the source frequency needs to be a lot larger than 10 GHz and the ring radius smaller than $10^2$ nm. These numbers of the values are realizable in future experiments.

Note that our discovery method actually utilises the limitation of the conformal invariance.
At the quantum critical point, the theory is scale-free and so is expected to be conformally invariant.
At zero temperature, the two-point functions (Green's function) of the CFT on a ring is completely determined by the conformal invariance with the dimensions of the operator, thus the response to the source to the first order is also determined from the symmetry; there is no room for the peculiarity of the SEM against the non-SEM to appear.
However, when this CFT on a ring
is put at a nonzero temperature, the conformal symmetry is reduced and cannot determine the correlator completely,\footnote{
For the determination of the retarded Green's function of any CFT on a line at nonzero temperature and its coincidence with the holographic gravity calculation, see \cite{Son:2002sd}. In general, when the system is put on a ring at nonzero temperature, the CFT partition function is a torus partition function which is not determined completely by the conformal invariance only. So the two-point functions which we study in this paper is not determined either.}
where the difference between the SEM and the non-SEM appears.
Only the SEM keeps the zero-temperature correlator intact in $T<1/a$, while the non-SEM correlator largely depends on temperature and is characterized by
the thermal free correlator.
It is an interesting observation that our discovery channel actually owes to the fact that experiments cannot reach the exact zero temperature; at zero temperature any system is just dictated by the whole conformal symmetry and the SEMs are not well-distinguishable.

Finally let us argue the universality of the SEM model we have adopted in this paper.
Our SEM model is the simplest choice, a bulk massless scalar field theory on the two different background geometries, the BTZ black hole and the AdS soliton.
Here, we present some arguments for the universality of our results against the following typical modifications/generalizations of the SEM theory.

\begin{itemize}
    \item Addition of other fields, different spins.
    
When there are several orders in the material, or there are orders associated with global symmetries or magnetization, it is natural to expect several operators with different spins are involved in theoretical realization of the ring-shaped material. In the AdS/CFT dictionary, those situations correspond to having several bulk fields with different spins. Will these changes affect our theoretical results?
First, our geometrical optics approximation works well while is applied to massless higher spin fields. Second, 
our calculation of the response function is robust against the addition of scalar fields, because it needs only Green's functions on the ring or in the bulk. 
Even when there are several fields, one can diagonalize the Hamiltonian and apply the Green function method for each diagonalized component. 
Therefore we expect that our basic strategy works for any model with small field amplitudes, and our results are robust.

    \item Change of the mass of the field. 

One can generalize the SEM model by adding a mass to the bulk field. According to the AdS/CFT dictionary it changes the scaling dimension of the boundary operator $O$ that couples to the source.
The masslessness of the field is nothing special, since the mass squared of the bulk field can be negative and just needs to be larger than or equal to the Breitenlohner-Freedman bound. Thus it is expected that the change of the mass of the bulk field will not give any drastic change of our result.
Additionally, high energy modes tend to be dominant in boundary-to-boundary propagators, and the mass can effectively be ignored.
Thus, we can expect that similar results will appear in massive cases.
Also note that, as \eqref{eq: imaging} is just a mathematical operation (a class of finite Fourier transforms), it can also be applied formally to any $O$, not limited to light waves.

    \item Deformation of the background curved geometries.

The low temperature phase of our SEM is described by the AdS soliton geometry.
Although this geometry is ensured by the conformal symmetry under the holographic principle and thus does not allow any deformation,\footnote{
In the AdS/CFT correspondence, the CFT$_2$ on $\mathbb R^{1,1}$ in the ground state corresponds to the pure Poincar\'e AdS$_3$.
This is because the conformal symmetry of the CFT$_2$ is SO(2,2) which is the isometry of the bulk geometry.
Now, when the CFT$_2$ is put on a compactified ring, the conformal symmetry is SO(2,2)$/\Gamma$, where $\Gamma$ is the discrete spatial translation by the circumference $a$.
In the gravity side, SO(2,2)$/\Gamma$ still guarantees the same number of the Killing vectors as that of maximally symmetric spacetimes, and hence the geometry must be locally AdS$_3$.
There are two possible spacetimes which possess SO(2,2)$/\Gamma$; one is the AdS soliton, which we considered, and the other is obtained simply by compactifying the Poincar\'e AdS$_3$ with $\Gamma$.
Only the former spacetime holographically reproduces the entanglement entropy of the CFT$_2$ on $\mathbb R^1\times\mathbb S^1$, and the one which is favored in the viewpoint of the free energy computed from the Euclidean Einstein gravity, is also the former.
Therefore, this symmetry argument signals out only the AdS soliton geometry as the gravitational dual of the CFT$_2$ on the ring.} 
realistic experiments may not be performed exactly
on the quantum critical point and lead to possible deformations of the 
AdS soliton geometry.
In particular, our discovery scheme of SEMs is owing strongly to the fact that all null geodesics focus on the unique antipodal point on the ring, which will not be exact in the deformed spacetimes.
However, null geodesics should still be focused near the antipodal point, and hence there will remain the fundamental feature that the image at the antipodal point is far stronger than those at different points.

Another issue is the interference of waves. When the focus of null geodesics is blurred due to the change of the geometry, the image at the antipodal point in the low temperature phase may acquire nodes due to the interference, while there is no node in the image of the exact AdS soliton case.
The nodal image would be similar to that of the non-SEM model with $m\lesssim \omega$, such as $m = 45\pi/a$ in Fig.\ref{fig: transformed image}, and hence it could falsify our discrimination method of the emergent spacetime.
But, we can argue that the $\theta_\mathrm{obs}$-dependence of the image cures our method.
It is seen in the figure \ref{fig: transformed image} that the magnitude of the image of $m = 45\pi/a$ does not vary so much when we change the value of $\theta_\mathrm{obs}$.
On the other hand, even when the AdS soliton is deformed, the image in the low temperature phase will vary significantly in $\theta_\mathrm{obs}$, because null geodesics do not reach boundary points not near the antipodal point, as described above. Therefore we expect that our method is sustainable against the continuous deformation of the background geometry of the emergent spacetime.

\end{itemize}

With these expectations, we hope that experiments are performed in the near future to find a spacetime-emergent material, for quantum gravity experiments.

\section*{Acknowledgement}
We would like to thank Yoichi Yanase for his valuable advices, and Keiju Murata for discussions. The work of K.H.\ is supported in part by JSPS KAKENHI Grant Number JP22H05115, JP22H05111 and JP22H01217.
The work of D.T.\ is supported by Grant-in-Aid for JSPS Fellows No.\ 22J20722.
The work of K.T.\ is supported by by MEXT Q-LEAP (Grant No. JPMXS0118067634) and by JSPS Grant-in-Aid for Scientific Research (S) (Grant No. JP21H05017).
The work of S.Y.\ is supported in part by JSPS KAKENHI Grant Numbers 22H04473 and 22H01168.

\appendix

\section{Plot of response functions}
\label{app:A}

\begin{figure}
\centering
	\begin{tabular}{cc}
		\begin{minipage}{0.5\columnwidth}
            \centering
			\includegraphics[height = 5.4cm]{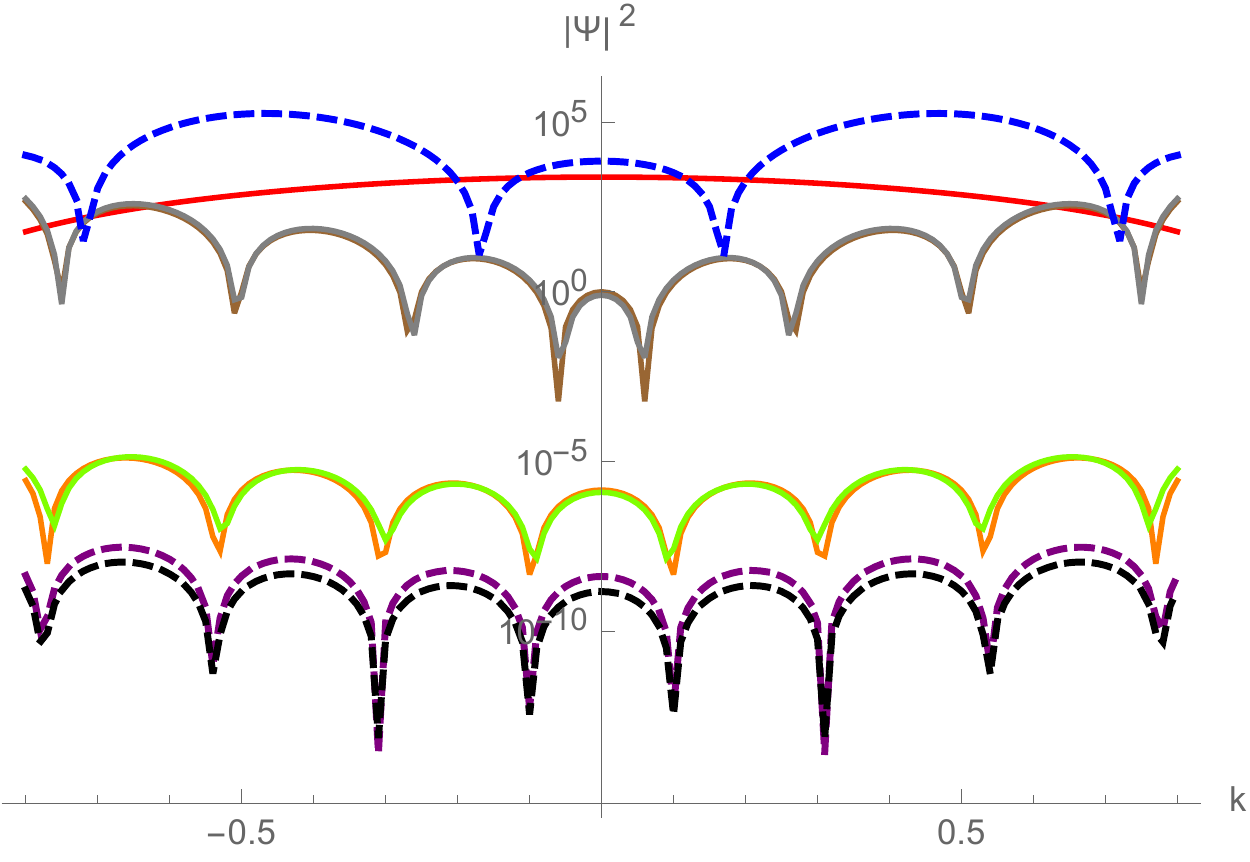}
		\end{minipage}&
		
		\begin{minipage}{0.5\columnwidth}
            \centering
			\includegraphics[height = 5.4cm]{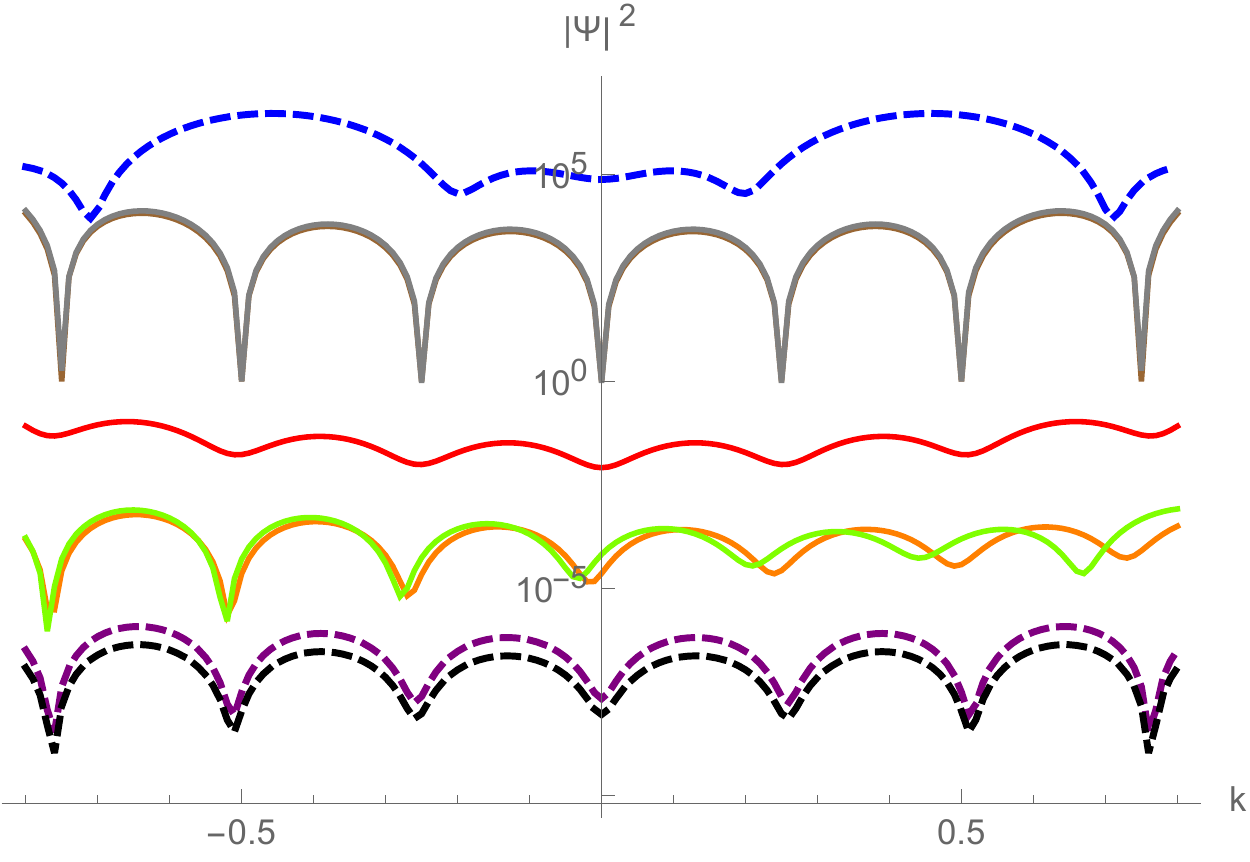}
		\end{minipage}\\
        $\theta_\mathrm{obs}=a/2$ & $\theta_\mathrm{obs}=a/3$
	\end{tabular}\vspace{36pt}
		\includegraphics[height = 5.3cm]{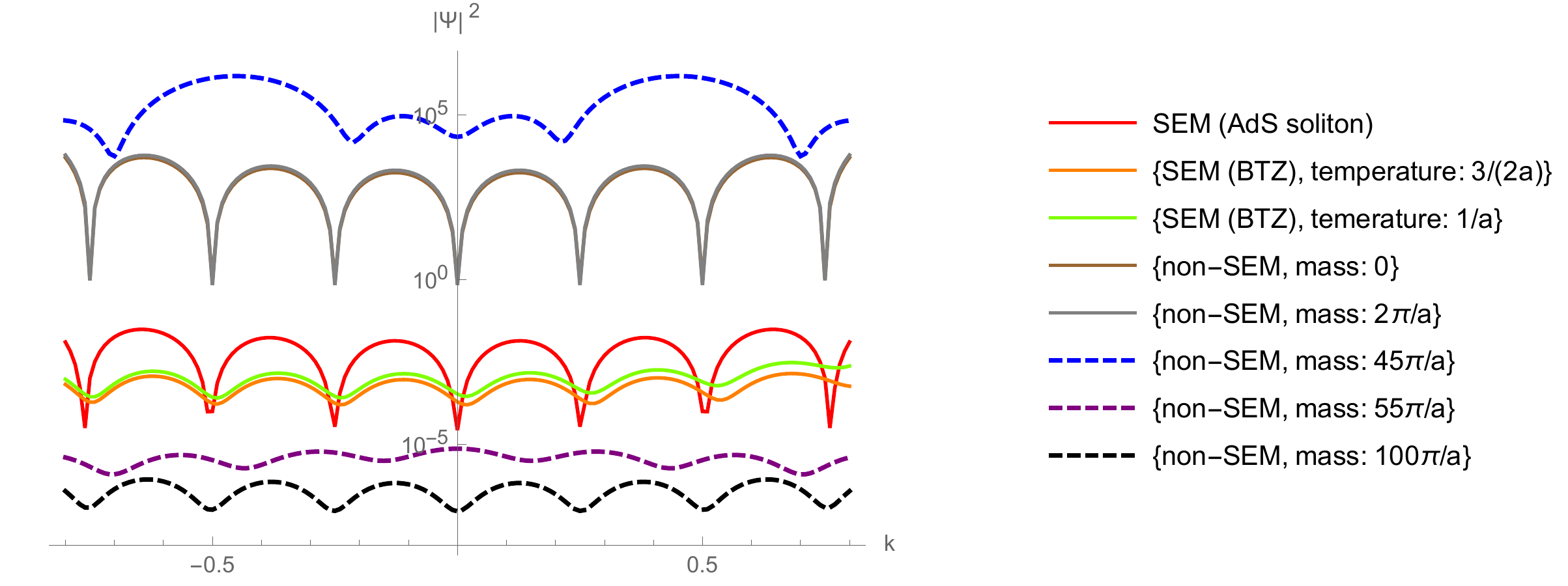}\\
        $\theta_\mathrm{obs}=a/4$\hspace{160pt}~
	\caption{
	The plots of the images of the response functions.
	}
 \label{fig: all responses}
\end{figure}

Here in this appendix, we show details of the response functions and the images. First, in Fig.~\ref{fig: all responses}
we present the detailed images for various non-SEM and SEM models.
	Since we did not care about the overall factors, we cannot compare the magnitude difference between the two models (comparison of relative magnitude within the same model is still accurate).

\begin{figure}
    \centering
    	
    \includegraphics[width = 15cm]{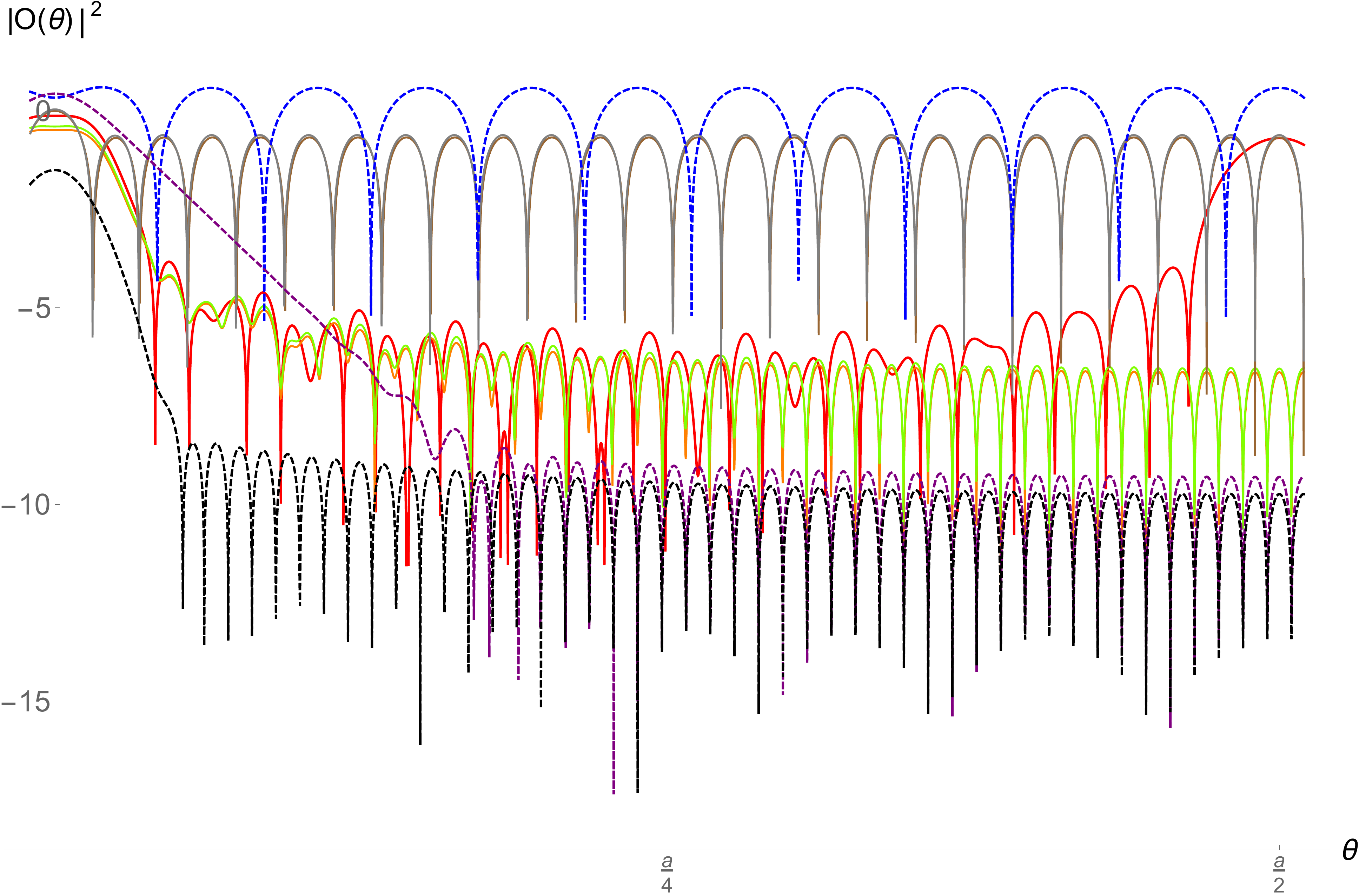}
    \caption{Response functions. The legends are shared with Fig.\ref{fig: all responses} (the AdS soliton case is shown by the red line).
    }
    \label{fig: response_raw}
\end{figure}

We also present the plot of the response functions which are to be Fourier-transformed, for readers' reference. See Fig.\ref{fig: response_raw}. The legends of the plot is shared with Fig.\ref{fig: all responses}.
The relative overall normalization between the two models should not be trusted, as was explained above.

Among the plots, the response function of the AdS soliton case
(which is the SEM case at the low temperature phase) exhibits the unique feature that the amplitude of the response function grows by several orders at the antipodal point of the ring.
This feature is consistent with our imaging.\footnote{Note that the imaging was necessary to probe the bulk geometry directly,
because the response function itself is subject to phase interference which needs to be decoded by the Frourier transform.}

\bibliographystyle{jhep} 
\bibliography{ref}

\providecommand{\href}[2]{#2}\begingroup\raggedright\begin{thebibliography}{10}

\bibitem{Maldacena:1997re}
J.M.~Maldacena, \emph{{The Large N limit of superconformal field theories and
  supergravity}}, \href{https://doi.org/10.1023/A:1026654312961}{\emph{Adv.
  Theor. Math. Phys.} {\bfseries 2} (1998) 231}
  [\href{https://arxiv.org/abs/hep-th/9711200}{{\ttfamily hep-th/9711200}}].

\bibitem{pelat2020acoustic}
A.~Pelat, F.~Gautier, S.C.~Conlon and F.~Semperlotti, \emph{The acoustic black
  hole: A review of theory and applications}, {\emph{Journal of Sound and
  Vibration} {\bfseries 476} (2020) 115316}.

\bibitem{soluyanov2015type}
A.A.~Soluyanov, D.~Gresch, Z.~Wang, Q.~Wu, M.~Troyer, X.~Dai et~al.,
  \emph{Type-ii weyl semimetals}, {\emph{Nature} {\bfseries 527} (2015) 495}.

\bibitem{Karch:2007pd}
A.~Karch and A.~O'Bannon, \emph{{Metallic AdS/CFT}},
  \href{https://doi.org/10.1088/1126-6708/2007/09/024}{\emph{JHEP} {\bfseries
  09} (2007) 024} [\href{https://arxiv.org/abs/0705.3870}{{\ttfamily
  0705.3870}}].

\bibitem{Hartnoll:2008vx}
S.A.~Hartnoll, C.P.~Herzog and G.T.~Horowitz, \emph{{Building a Holographic
  Superconductor}},
  \href{https://doi.org/10.1103/PhysRevLett.101.031601}{\emph{Phys. Rev. Lett.}
  {\bfseries 101} (2008) 031601}
  [\href{https://arxiv.org/abs/0803.3295}{{\ttfamily 0803.3295}}].

\bibitem{Hartnoll:2008kx}
S.A.~Hartnoll, C.P.~Herzog and G.T.~Horowitz, \emph{{Holographic
  Superconductors}},
  \href{https://doi.org/10.1088/1126-6708/2008/12/015}{\emph{JHEP} {\bfseries
  12} (2008) 015} [\href{https://arxiv.org/abs/0810.1563}{{\ttfamily
  0810.1563}}].

\bibitem{Hashimoto:2018okj}
K.~Hashimoto, S.~Kinoshita and K.~Murata, \emph{{Imaging black holes through
  the AdS/CFT correspondence}},
  \href{https://doi.org/10.1103/PhysRevD.101.066018}{\emph{Phys. Rev. D}
  {\bfseries 101} (2020) 066018}
  [\href{https://arxiv.org/abs/1811.12617}{{\ttfamily 1811.12617}}].

\bibitem{Hashimoto:2019jmw}
K.~Hashimoto, S.~Kinoshita and K.~Murata, \emph{{Einstein Rings in
  Holography}},
  \href{https://doi.org/10.1103/PhysRevLett.123.031602}{\emph{Phys. Rev. Lett.}
  {\bfseries 123} (2019) 031602}
  [\href{https://arxiv.org/abs/1906.09113}{{\ttfamily 1906.09113}}].

\bibitem{Kaku:2021xqp}
Y.~Kaku, K.~Murata and J.~Tsujimura, \emph{{Observing black holes through
  superconductors}}, \href{https://doi.org/10.1007/JHEP09(2021)138}{\emph{JHEP}
  {\bfseries 09} (2021) 138}
  [\href{https://arxiv.org/abs/2106.00304}{{\ttfamily 2106.00304}}].

\bibitem{Zeng:2022woh}
X.-X.~Zeng, H.~Zhang and W.-L.~Zhang, \emph{{Holographic Einstein Ring of a
  Charged AdS Black Hole}},  \href{https://arxiv.org/abs/2201.03161}{{\ttfamily
  2201.03161}}.

\bibitem{Gubser:1998bc}
S.S.~Gubser, I.R.~Klebanov and A.M.~Polyakov, \emph{{Gauge theory correlators
  from noncritical string theory}},
  \href{https://doi.org/10.1016/S0370-2693(98)00377-3}{\emph{Phys. Lett. B}
  {\bfseries 428} (1998) 105}
  [\href{https://arxiv.org/abs/hep-th/9802109}{{\ttfamily hep-th/9802109}}].

\bibitem{Witten:1998qj}
E.~Witten, \emph{{Anti-de Sitter space and holography}},
  \href{https://doi.org/10.4310/ATMP.1998.v2.n2.a2}{\emph{Adv. Theor. Math.
  Phys.} {\bfseries 2} (1998) 253}
  [\href{https://arxiv.org/abs/hep-th/9802150}{{\ttfamily hep-th/9802150}}].

\bibitem{Klebanov:1999tb}
I.R.~Klebanov and E.~Witten, \emph{{AdS / CFT correspondence and symmetry
  breaking}}, \href{https://doi.org/10.1016/S0550-3213(99)00387-9}{\emph{Nucl.
  Phys. B} {\bfseries 556} (1999) 89}
  [\href{https://arxiv.org/abs/hep-th/9905104}{{\ttfamily hep-th/9905104}}].

\bibitem{matsumoto2002magnon}
M.~Matsumoto, B.~Normand, T.~Rice and M.~Sigrist, \emph{Magnon dispersion in
  the field-induced magnetically ordered phase of {TlCuCl$_3$}}, {\emph{Phys.
  Rev. Lett.} {\bfseries 89} (2002) 077203}.

\bibitem{Matsumoto2004FieldAP}
M.~Matsumoto, B.~Normand, T.M.~Rice and M.~Sigrist, \emph{Field- and
  pressure-induced magnetic quantum phase transitions in {TlCuCl$_3$}},
  {\emph{Phys. Rev. B} {\bfseries 69} (2004) 054423}.

\bibitem{sachdev2011quantum}
S.~Sachdev and B.~Keimer, \emph{Quantum criticality}, {\emph{Physics Today}
  {\bfseries 64} (2011) 29}.

\bibitem{cavadini2001magnetic}
N.~Cavadini, G.~Heigold, W.~Henggeler, A.~Furrer, H.-U.~G{\"u}del,
  K.~Kr{\"a}mer et~al., \emph{Magnetic excitations in the quantum spin system
  {TlCuCl$_3$}}, {\emph{Physical Review B} {\bfseries 63} (2001) 172414}.

\bibitem{tanaka2001observation}
H.~Tanaka, A.~Oosawa, T.~Kato, H.~Uekusa, Y.~Ohashi, K.~Kakurai et~al.,
  \emph{Observation of field-induced transverse n{\'e}el ordering in the spin
  gap system {TlCuCl$_3$}}, {\emph{Journal of the Physical Society of Japan}
  {\bfseries 70} (2001) 939}.

\bibitem{Son:2002sd}
D.T.~Son and A.O.~Starinets, \emph{{Minkowski space correlators in AdS / CFT
  correspondence: Recipe and applications}},
  \href{https://doi.org/10.1088/1126-6708/2002/09/042}{\emph{JHEP} {\bfseries
  09} (2002) 042} [\href{https://arxiv.org/abs/hep-th/0205051}{{\ttfamily
  hep-th/0205051}}].

\end{thebibliography}\endgroup
\end{document}